\documentclass[10pt,reqno,a4paper,final]{amsart}
\usepackage[british]{babel}
\usepackage{amssymb,amstext,amsthm,eucal,bbm,amscd,bbold}
\usepackage[margin=1in]{geometry}
\usepackage{array}
\usepackage[svgnames,table]{xcolor}
\usepackage[T1]{fontenc}
\usepackage[utf8]{inputenc}
\usepackage{mathrsfs}
\usepackage{verbatim}
\usepackage{pdfsync}
\usepackage[final]{microtype}
\usepackage[nosort]{cite}
\usepackage{adjustbox}
\usepackage{bm}
\usepackage{booktabs} 
\usepackage{caption}
\usepackage{appendix}
\usepackage{chngcntr}
\usepackage{apptools}
\usepackage{romanbar}
\usepackage{enumitem}
\usepackage{slashed}

\usepackage{tikz,pgfplots}
\pgfplotsset{compat=1.18}
\usepgfplotslibrary{fillbetween}
\usetikzlibrary{calc,arrows,arrows.meta,cd,shapes.geometric,positioning,through,decorations.markings,decorations.text}
\tikzset{>=latex}
\usepackage[unicode,pagebackref=true]{hyperref}
\hypersetup{%
  pdftitle   = {Galilei particles revisited},
  pdfkeywords = {Galilei, Bargmann, coadjoint orbits, particles},
  pdfauthor  = {José Figueroa-O'Farrill, Simon Pekar, Alfredo Pérez,
    Stefan Prohazka},
  pdfcreator = {\LaTeX\ with package \flqq hyperref\frqq},
  linkcolor=NavyBlue,
  citecolor=ForestGreen,
  urlcolor=OrangeRed,
  anchorcolor=OrangeRed,
  colorlinks=true}
\renewcommand*{\backref}[1]{}
\renewcommand*{\backrefalt}[4]{%
  \ifcase #1%
  \or [Page~#2.]%
  \else [Pages~#2.]%
  \fi%
}
\theoremstyle{plain}

\theoremstyle{definition}

\newcommand{\g}{\mathfrak{g}}

\newcommand{\fk}{\mathfrak{k}}

\renewcommand{\c}{\mathfrak{c}}
\renewcommand{\d}{\partial}
\newcommand{\gl}{\mathfrak{gl}}

\newcommand{\so}{\mathfrak{so}}

\newcommand{\ft}{\mathfrak{t}}
\newcommand{\fm}{\mathfrak{m}}

\newcommand{\be}{\boldsymbol{e}}
\newcommand{\x}{\boldsymbol{x}}
\newcommand{\y}{\boldsymbol{y}}

\newcommand{\bk}{\boldsymbol{k}}
\newcommand{\bv}{\boldsymbol{v}}
\newcommand{\bj}{\boldsymbol{j}}

\newcommand{\p}{\boldsymbol{p}}
\newcommand{\ba}{\boldsymbol{a}}
\newcommand{\bb}{\boldsymbol{b}}

\newcommand{\bzero}{\boldsymbol{0}}
\newcommand{\balpha}{\boldsymbol{\alpha}}
\newcommand{\bbeta}{\boldsymbol{\beta}}
\newcommand{\bkappa}{\boldsymbol{\kappa}}
\newcommand{\bP}{\boldsymbol{P}}
\newcommand{\bB}{\boldsymbol{B}}
\newcommand{\bpi}{\boldsymbol{\pi}}

\newcommand{\sg}{\boldsymbol{\mathsf{g}}}

\newcommand{\sA}{\boldsymbol{\mathsf{A}}}
\newcommand{\sM}{\boldsymbol{\mathsf{M}}}

\newcommand{\cO}{\mathscr{O}}

\newcommand{\eH}{\mathcal{H}}

\newcommand{\eO}{\mathcal{O}}

\newcommand{\Cl}{C\ell}

\renewcommand{\Re}{\operatorname{Re}}
\renewcommand{\Im}{\operatorname{Im}}
\newcommand{\ad}{\operatorname{ad}}
\newcommand{\Ort}{\operatorname{O}}

\newcommand{\Ad}{\operatorname{Ad}}

\newcommand{\dvol}{\operatorname{dvol}}

\newcommand{\Tr}{\operatorname{Tr}}

\newcommand{\CP}{\mathbb{CP}}

\newcommand{\RR}{\mathbb{R}}

\newcommand{\ZZ}{\mathbb{Z}}

\newcommand{\CC}{\mathbb{C}}

\newcommand{\GL}{\operatorname{GL}}
\newcommand{\ISO}{\operatorname{ISO}}
\newcommand{\SO}{\operatorname{SO}}
\newcommand{\SU}{\operatorname{SU}}
\newcommand{\U}{\operatorname{U}}

\newcommand{\Spin}{\operatorname{Spin}}

\newcommand{\zbar}{\overline{z}}

\definecolor{dkgr}{rgb}{0,0.6,0}
\definecolor{gris}{rgb}{0.5,0.5,0.5}

\providecommand*{\vd}{\delta}
\renewcommand*{\vd}{\delta}

\allowdisplaybreaks[0]
\numberwithin{equation}{section}
\AtAppendix{\counterwithin{lemma}{section}}

\begin{document}

\title{Galilei particles revisited}

\author[Figueroa-O'Farrill]{José Miguel Figueroa-O'Farrill}
\author[Pekar]{Simon Pekar}
\author[Pérez]{Alfredo Pérez}
\author[Prohazka]{Stefan Prohazka}
\address[JMF]{Maxwell Institute and School of Mathematics, The University
  of Edinburgh, James Clerk Maxwell Building, Peter Guthrie Tait Road,
  Edinburgh EH9 3FD, Scotland, United Kingdom}
\address[SiP]{Centre de Physique Théorique – CPHT, École polytechnique, CNRS, Institut Polytechnique de Paris, 91120 Palaiseau Cedex, France}
\address[AP]{Centro de Estudios Científicos (CECs), Avenida Arturo Prat 514, Valdivia, Chile}
\address[AP]{Facultad de Ingeniería, Arquitectura y Diseño, Universidad San Sebastián, sede Valdivia, General Lagos 1163, Valdivia 5110693, Chile}
\address[StP]{University of Vienna, Faculty of Physics, Mathematical Physics, Boltzmanngasse 5, 1090, Vienna, Austria}
\email[JMF]{\href{mailto:j.m.figueroa@ed.ac.uk}{j.m.figueroa@ed.ac.uk}, ORCID: \href{https://orcid.org/0000-0002-9308-9360}{0000-0002-9308-9360}}
\email[AP]{\href{mailto:alfredo.perez@uss.cl}{alfredo.perez@uss.cl}, ORCID: \href{https://orcid.org/0000-0003-0989-9959}{0000-0003-0989-9959}}
\email[SiP]{\href{mailto:simon.pekar@polytechnique.edu}{simon.pekar@polytechnique.edu},
  ORCID: \href{https://orcid.org/0000-0002-0765-8986}{0000-0002-0765-8986}}
\email[StP]{\href{mailto:stefan.prohazka@univie.ac.at}{stefan.prohazka@univie.ac.at},
  ORCID: \href{https://orcid.org/0000-0002-3925-3983}{0000-0002-3925-3983}}

\begin{abstract}
We revisit the classifications of classical and quantum galilean particles: that is, we fully classify homogeneous symplectic manifolds and unitary irreducible projective representations of the Galilei group. Equivalently, these are coadjoint orbits and unitary irreducible representations of the Bargmann group, the universal central extension of the Galilei group. We provide an action principle in each case, discuss the nonrelativistic limit, as well as exhibit, whenever possible, the unitary irreducible representations in terms of fields on Galilei spacetime. Motivated by a forthcoming study of planons we pay close attention to the mobility of the less familiar massless Galilei particles.
\end{abstract}

\maketitle
\tableofcontents

\section{Introduction}
\label{sec:introduction}

Galilei, Minkowski and Carroll
spacetimes~\cite{Bacry:1968zf,Figueroa-OFarrill:2018ilb} are
distinguished Klein models for galilean, lorentzian and carrollian
Cartan geometries, respectively. They are affine spaces admitting a
transitive action of the Galilei, Poincaré and Carroll groups,
respectively, which play the rôle of relativity groups for these
spacetimes. The latter are the arena for both particle dynamics
(classical and quantum) as well as quantum field theories: galilean,
(the misnamed) relativistic and carrollian, respectively. Despite the
physical spacetimes being the arena for dynamics, the actual degrees
of freedom are often described by other homogeneous spaces of the
groups in question: homogeneous symplectic manifolds in the case of
classical particle dynamics, and momentum orbits in the case of
quantum particles and fields.

These relativity groups are semidirect products $G = K \ltimes T$,
where $T$ is abelian (the translations) and $K$ consists, roughly, of
rotations and boosts. Of these relativity groups, the best understood
is arguably the Poincaré group, where $K$ is the Lorentz group. The
Poincaré group has played a starring rôle in Physics for more than a
century: it underlies Special Relativity and Relativistic Quantum
Field Theory, and its associated Cartan geometry (i.e., lorentzian
geometry) underlies General Relativity. Classical Poincaré particles
correspond to coadjoint orbits of the Poincaré group and they were
first classified by Arens in \cite{MR0299100}. Its unitary irreducible
representations were famously classified by Wigner \cite{MR1503456},
pioneering the method of induced representations which would find its
most general expression in Mackey's theory \cite{MR0396826}.

By contrast, the Carroll group is more recent. Introduced by
Lévy-Leblond in \cite{Levy1965} and independently by Sen~Gupta in
\cite{SenGupta1966OnAA}, its coadjoint orbits were studied initially
in the Appendix of \cite{Duval:2014uoa} and more recently also in
\cite{Figueroa-OFarrill:2023vbj} in the context of fractons. The
unitary irreducible representations of the Carroll group were recently
determined in \cite{Figueroa-OFarrill:2023qty} using the method of
induced representations (see also~\cite{Levy1965,deBoer:2021jej}).

The Galilei group is of course the oldest of these relativity groups
and it is the subject of this paper.  In contrast to both the Poincaré
and Carroll groups, the Galilei group has nontrivial symplectic
cohomology in the language of Souriau \cite{MR0260238}. This means
that its homogeneous symplectic manifolds are not all coadjoint orbits
of the Galilei group, but of its universal central extension: the
eponymous group introduced by Bargmann in \cite{MR0058601} in a
quantum mechanical context.  In that paper, Bargmann showed that the
unitary irreducible ray representations of the Galilei group are
honest unitary irreducible representations of the Bargmann group, but
stopped short of their classification.  The coadjoint orbits of the
Bargmann group were initially studied by Souriau \cite[§14]{MR0260238}
and are also discussed by Guillemin and Sternberg in
\cite[§54]{MR1066693}, but to the best of our knowledge a full
analysis of the resulting particle dynamics has not been done before.
The story of the unitary irreducible representations is more tortuous.
It was Inönü and Wigner \cite{MR0050594} who first considered the
unitary irreducible representations of the Galilei group.  In that
paper they consciously restricted to honest representations of the
connected Galilei group, thus missing massive representations, and
moreover they did not consider the simply-connected cover of the
Galilei group, thus missing representations with half-integer
spin. Brennich \cite{MR0275770} classified the unitary irreducible ray
representations of the simply-connected Galilei group and also of its
extension by parity and time reversal automorphisms.  His labeling of
representations is somewhat redundant, as we shall see.  The unitary
irreducible ray representations were also determined by Lévy-Leblond in
\cite{MR0466427} with one minor omission, as we shall discuss below.
These works pay no attention to identifying the homogeneous vector
bundles whose sections carry the representations.

In recent years novel quasiparticles with the distinctive property of
having only restricted mobility (see
\cite{Nandkishore:2018sel,Pretko:2020cko,Grosvenor:2021hkn} for
reviews) have challenged, and hence advanced, our understanding of
conventional quantum field
theories~\cite{Cordova:2022ruw,Brauner:2022rvf}. One way to understand
them~\cite{Pretko:2016kxt,Pretko:2016lgv} is coming from theories with
higher moment conservation laws like, e.g., dipole moment
conservation. These exotic symmetries can be studied systematically
and it was observed~\cite{Gromov:2018nbv} that the symmetries of
theories with conserved dipole and trace of the quadrupole moment
closely resemble the Bargmann algebra (similar to the relation between
fractons and Carroll
particles~\cite{Bidussi:2021nmp,Marsot:2022imf,Figueroa-OFarrill:2023vbj,Figueroa-OFarrill:2023qty}).
It is then natural to ask if and how their particles are related,
which leads to study the classical and quantum elementary and
composite systems. These particles indeed play a rôle in applications
(e.g.,~\cite{Perez:2023uwt}) and to be able to contrast the planons
with the Galilei particles~\cite{Planons:2024} we find it useful to
review the Galilei particles independently.

We make no strong claims of originality, but we are unaware of a
resource that uniformly covers the topics of this article at the same
level of completeness.  We therefore think it may be useful to collect
these results in a uniform way, using contemporary language and paying
close attention to the geometrical nature of the representations. In
addition we discuss the particle dynamics associated to the different
coadjoint orbits as well as, whenever possible, a description of the
unitary irreducible representations as fields in Galilei spacetime.

This paper is organised as follows. In
Section~\ref{sec:bargmann-group} we introduce the Bargmann group. We
do not define it as the universal central extension of the Galilei
group, but rather as the subgroup of the Poincaré group in one
dimension higher which stabilises a nonzero null translation generator
and only then show that its Lie algebra is a central extension of the
Galilei algebra. In Section~\ref{sec:adjo-coadj-repr} we discuss the
adjoint and coadjoint representations of the Bargmann group and in
Section~\ref{sec:maurer-cartan-one} we write explicit expressions for
the left-invariant Maurer--Cartan one-form on the Bargmann group, for
later use in deriving action functionals for classical Galilei
particles. Until this point we have been working in generic dimension,
but starting from Section~\ref{sec:case-n=3} we restrict our attention
to the case of four-dimensional Galilei spacetime. In
Section~\ref{sec:automorphisms} we determine for later use the
automorphisms of the Bargmann algebra which act trivially on the
rotational subalgebra and we give an expression for the group
automorphisms which integrate them. In
Section~\ref{sec:coadjoint-orbits-3d} we determine the coadjoint
orbits of the Bargmann group. They are summarised in
Table~\ref{tab:coadjoint-orbits}, which gives equations for each of
the orbits. We then discuss how this classification changes when we
extend the Bargmann group by parity and time reversal. Finally, we
discuss the geometric structure of the coadjoint orbits as bundles
over the momentum orbits, which is summarised in
Table~\ref{tab:structure-orbits}. In Section~\ref{sec:action-galil} we
study the particle actions associated to each of the coadjoint orbits
and determine the corresponding dynamics (with further details
delegated to Appendix~\ref{sec:symm-mass-spinl}). In
Section~\ref{sec:particle-dynamics} we discuss a group-theoretical
approach to particle dynamics and compare with the results in
Section~\ref{sec:action-galil}. In Section~\ref{sec:from-poinc-galil}
we contrast the energy-momentum orbits and the limit from Poincaré to
Galilei (see Figure~\ref{fig:momentum-orb}). In
Section~\ref{sec:unit-irred-repr} we classify the unitary irreducible
representations of the Bargmann group using the method of induced
representations. The results are summarised in
Table~\ref{tab:orbits-uirreps}. We compare our classification with
those of Inönü--Wigner \cite{MR0050594}, Brennich \cite{MR0275770} and
Lévy-Leblond \cite{MR0466427}. Finally, in
Section~\ref{sec:galil-field-theor} we discuss some realisations of
these unitary irreducible representations in terms of fields in
Galilei spacetime.

\section{The Bargmann group and its Lie algebra}
\label{sec:bargmann-group}

In this section we define the Bargmann group and work out its Lie
algebra.

The ($n+1$)-dimensional Bargmann group is the subgroup $G$ of the
($n+2$)-dimensional Poincaré group which leaves invariant a null
translation under the adjoint representation on its Lie algebra.  The
Poincaré group sits inside the affine group, which in turn embeds
inside the linear group in one higher dimension.  Therefore the
($n+1$)-dimensional Bargmann group sits naturally inside $\GL(n+3,\RR)$.

The ($n+2$)-dimensional Poincaré group, by which we mean the subgroup
of isometries of ($n+2$)-dimensional Minkowski spacetime, is the
subgroup of $\GL(n+3,\RR)$ given by the set of matrices
\begin{equation}
  \label{eq:poincare-inside-gl}
  \left\{
    \begin{pmatrix}
      L & a \\ 0 & 1
    \end{pmatrix} ~\middle |~ a \in \RR^{n+2},\quad L \in \GL(n+2,\RR) \quad\text{and}\quad L^T\eta L = \eta \right\},
\end{equation}
where
\begin{equation}
  \label{eq:minkowski-metric}
  \eta =
  \begin{pmatrix}
    0 & 1 & \bzero^T\\
    1 & 0 & \bzero^T\\
    \bzero & \bzero & I_n
  \end{pmatrix} \qquad\text{with $I_n$ the identity matrix of size $n$}.
\end{equation}
Notice that we have chosen a Witt frame $(\be_+,\be_-,\be_a)$ for the
lorentzian vector space $(\RR^{n+2},\eta)$, in such a way that
$\eta(\be_+,\be_-) = 1$, $\eta(\be_a, \be_b) = \delta_{ab}$ and all
other inner products vanish.  The Bargmann group is therefore given by
the set of matrices
\begin{equation}
  \label{eq:bargmann-inside-gl}
  \left\{
    \begin{pmatrix}
      L & a \\ 0 & 1
    \end{pmatrix} ~\middle |~ a \in \RR^{n+2},\quad L \in
    \GL(n+2,\RR),\quad L^T\eta L = \eta\quad\text{and}\quad L\be_+ = \be_+ \right\}.
\end{equation}
The matrices $L \in \GL(n+2,\RR)$ in the Bargmann group can be seen to
take the following form
\begin{equation}
  \label{eq:homog-barg}
  L =
  \begin{pmatrix}
    1 & -\tfrac12 \|\bv\|^2 & \bv^TR\\
    0 & 1 & \bzero^T\\
    \bzero & -\bv & R
  \end{pmatrix}\qquad\text{where $R \in \Ort(n)$ and $\bv \in \RR^n$.}
\end{equation}
The subgroup $H \subset \GL(n+2,\RR)$ consisting of such matrices is
isomorphic to the euclidean group $\RR^n \rtimes \Ort(n)$.  In
other words, if is diffeomorphic to $\RR^n \times \Ort(n)$ where
the element of $H$ corresponding to $(\bv,R) \in \RR^n \times
\Ort(n)$ is given by the matrix $L$ in
equation~\eqref{eq:homog-barg}.  That element, which we
denote\footnote{We use the notation $\sg(\ldots)$, $\sA(\ldots)$ and
  $\sM(\ldots)$ to represent group elements, Lie algebra elements and
  elements of the dual of the Lie algebra (``moments'') parametrised
  by the data inside the parentheses.} by
$\sg(\bv,R)$, can be factorised as
\begin{equation}
  \label{eq:homog-barg-factor}
  \begin{pmatrix}
    1 & -\tfrac12 \|\bv\|^2 & \bv^TR\\
    0 & 1 & \bzero^T\\
    \bzero & -\bv & R
  \end{pmatrix} =
  \begin{pmatrix}
    1 & -\tfrac12 \|\bv\|^2 & \bv^T\\
    0 & 1 & \bzero^T\\
    \bzero & -\bv & I_n
  \end{pmatrix}
    \begin{pmatrix}
    1 & 0 & \bzero^T\\
    0 & 1 & \bzero^T\\
    \bzero & \bzero & R
  \end{pmatrix},  
\end{equation}
corresponding to the multiplication on $\RR^n \times \Ort(n)$
defined by
\begin{equation}
  \label{eq:homog-barg-mult}
  \sg(\bv_1, R_1) \sg(\bv_2,R_2) = \sg(\bv_1 + R_1 \bv_2, R_1 R_2).
\end{equation}

The Bargmann subgroup $G \subset \GL(n+3,\RR)$ thus consists of
matrices of the form
\begin{equation}
  \label{eq:explicit-barg-in-gl}
  \begin{pmatrix}
    1 & -\tfrac12 \|\bv\|^2 & \bv^TR & a_+\\
    0 & 1 & \bzero^T & a_-\\
    \bzero & -\bv & R & \ba\\
    0 & 0 & \bzero^T & 1\\
  \end{pmatrix},
\end{equation}
which factorises as
\begin{equation}
  \label{eq:barg-factor}
  \begin{pmatrix}
    1 & -\tfrac12 \|\bv\|^2 & \bv^TR & a_+\\
    0 & 1 & \bzero^T & a_-\\
    \bzero & -\bv & R & \ba\\
    0 & 0 & \bzero^T & 1\\
  \end{pmatrix} =
  \begin{pmatrix}
    1 & 0 & \bzero^T & a_+\\
    0 & 1 & \bzero^T & a_-\\
    \bzero & \bzero & I_n & \ba\\
    0 & 0 & \bzero^T & 1\\
  \end{pmatrix}
    \begin{pmatrix}
    1 & -\tfrac12 \|\bv\|^2 & \bv^T & 0\\
    0 & 1 & \bzero^T & 0\\
    \bzero & -\bv & I_n & \bzero\\
    0 & 0 & \bzero^T & 1\\
  \end{pmatrix}
  \begin{pmatrix}
    1 & 0 & \bzero^T & 0\\
    0 & 1 & \bzero^T & 0\\
    \bzero & \bzero & R & \bzero\\
    0 & 0 & \bzero^T & 1\\
  \end{pmatrix},
\end{equation}
corresponding to
\begin{equation}\label{eq:barg-element}
  \sg(a_+,a_-,\ba,\bv,R) = \sg(a_+,a_-,\ba,\bzero,I_n) \sg(0,0,\bzero,\bv,I_n)\sg(0,0,\bzero,\bzero,R).
\end{equation}
The Bargmann group is thus diffeomorphic to $\RR \times \RR \times
\RR^n \times \RR^n \times \Ort(n)$, with multiplication
defined by
\begin{multline}
  \label{eq:barg-mult}
  \sg(a_+,a_-,\ba,\bv,R) \sg(\alpha_+,\alpha_-,\balpha,\bbeta,\Sigma)\\ = \sg(a_+ + \alpha_+ +
  \bv\cdot R\balpha - \tfrac12 \alpha_- \|\bv\|^2, a_- + \alpha_-, \ba + R\balpha -
  \alpha_- \bv, \bv + R \bbeta, R\Sigma),
\end{multline}
from where we see that the identity element corresponds to $\sg(0,0,\bzero,\bzero,I_n)$.
From here we can work out the group inversion:
\begin{equation}
  \label{eq:barg-inv}
  \sg(a_+,a_-,\ba,\bv,R)^{-1} = \sg(-a_+ + \ba \cdot \bv + \tfrac12 a_-
  \|\bv\|^2, -a_-, -R^T(\ba + a_-\bv), -R^T\bv, R^T),
\end{equation}
where we have used that $R^T = R^{-1}$ in $\Ort(n)$.

The Lie algebra $\g$ of the Bargmann group embeds in $\gl(n+3,\RR)$
with image consisting of matrices of the form
\begin{equation}
  \label{eq:barg-alg-in-gl}
  \sA(x_+,x_-,\x, \y, X):=
  \begin{pmatrix}
    0 & 0 & \y^T & x_+\\
    0 & 0 & \bzero^T & x_-\\
    \bzero & -\y & X & \x\\
    0 & 0 & \bzero^T & 0
  \end{pmatrix}\qquad\text{where $X^T = - X$,}
\end{equation}
from where we can easily work out the Lie bracket:
\begin{multline}
  \label{eq:barg-alg-lie-bra}
  \left[ \sA(x^1_+, x^1_-, \x_1, \y_1, X_1), \sA(x^2_+, x^2_-, \x_2, \y_2,
    X_2)\right]\\
  = \sA( \y_1 \cdot \x_2 - \y_1 \cdot \x_1, 0, X_1 \x_2 -
  X_2 \x_1 + x^1_-\y_2 - x^2_-\y_1, X_1 \y_2 - X_2 \y_1 , [X_1,X_2]).
\end{multline}
If we introduce a basis $L_{ab}, B_a, P_a, H, M$ for the Bargmann
algebra in such a way that\footnote{The sign in the $-x_- H$ term is
  so that the Lie brackets are the ones we are familiar with.}
\begin{equation}
  \label{eq:barg-alg-basis}
   \sA(x_+,x_-,\x, \y, X) = x_+ M - x_- H + x^a P_a + y^a B_a + \tfrac12 X^{ab}L_{ab},
\end{equation}
we read off the following nonzero Lie brackets:
\begin{equation}
  \label{eq:barg-alg-lie-basis}
  \begin{split}
    [L_{ab},L_{cd}] &= \delta_{bc} L_{ad} - \delta_{ac} L_{bd} -  \delta_{bd} L_{ac} + \delta_{bd} L_{ac} \\
    [L_{ab}, B_c] &= \delta_{bc} B_a - \delta_{ac} B_b\\
    [L_{ab}, P_c] &= \delta_{bc} P_a - \delta_{ac} P_b\\
    [B_a, H] &= P_a \\
    [B_a, P_b] &= \delta_{ab} M,
  \end{split}
\end{equation}
which exhibits the Bargmann algebra as a central extension of the
Galilei algebra.  Indeed, the coadjoint orbits of the Bargmann
group coincide, up to covering, with the homogeneous symplectic
manifolds of the Galilei group.

The other homogeneous space of the Galilei group we shall be
interested in is Galilei spacetime itself.  It admits an effective
transitive action of the Galilei group and hence the Bargmann group
too acts transitively, but not effectively.  It is simple to describe
this action.  Let $M$ denote the Galilei spacetime: it is an affine
space diffeomorphic to $G/G_0$ with $G_0 = K \times Z$, where $Z$ is the
one-dimensional centre and $K$ is the homogeneous Galilei group.  Its
Lie algebra $\g_0$ is spanned by $L_{ab}, B_a, M$.  We choose a coset
representative $\zeta : M \to G$ defined by
$\zeta(t,\x) = \exp(t H + \x \cdot P)$.  The action of $G$ on $M$ is
induced by left-multiplication on $G$:
\begin{equation}
  \sg(a_+,a_-, \ba, \bv, R) \zeta(t,\x) = \zeta(t',\x') h
\end{equation}
for some $h \in G_0$, which depends in principle on
$a_+,a_-,\ba,\bv,R,t,\x$.  One can calculate the above product and
arrives at
\begin{equation}\label{eq:barg-on-gal-st}
\sg(a_+,a_-,\ba,\bv,R) : \begin{pmatrix} t \\ \x \end{pmatrix}
\mapsto
\begin{pmatrix}
  t' \\ \x'
\end{pmatrix} =
\begin{pmatrix}
  t + a_- \\
  R \x - t \bv + \ba
\end{pmatrix},
\end{equation}
from where we see that the central subgroup $Z$ acts trivially and
hence the action factors through the Galilei group $G/Z$.  The action
is via a sequence of affine transformations: rotation followed by a
galilean boost and followed by translations in both space and time.

\section{The adjoint and coadjoint representations}
\label{sec:adjo-coadj-repr}

The adjoint representation of $G$ on $\g$ is the derivative of the
conjugation action of $G$ on itself at the identity. Conjugation is
easily worked out from the formulae for multiplication
\eqref{eq:barg-mult} and inversion \eqref{eq:barg-inv}.  We find that
\begin{equation}
  \sg(a_+,a_-, \ba, \bv, R) \sg(\alpha_+, \alpha_-, \balpha,\bbeta, \Sigma)
  \sg(a_+, a_-, \ba, \bv, R)^{-1} = \sg(\alpha'_+, \alpha'_-, \balpha',\bbeta', \Sigma'),
\end{equation}
where
\begin{equation}
  \label{eq:barg-grp-conj}
  \begin{split}
    \alpha'_+ &= \alpha_+ + \ba \cdot \bv - \tfrac12 \alpha_-
    \|\bv\|^2 + \tfrac12 a_- (\|\bv\|^2 + \|\bv + R\bbeta\|^2) + \bv
    \cdot R\balpha - (\bv + R\bbeta)\cdot R \Sigma R^T(\ba + a_-
    \bv)\\
    \alpha'_- &= \alpha_-\\
    \balpha' &= R \balpha + \ba - \alpha_- \bv + a_- \bv + a_- R \bbeta - R\Sigma R^T \ba - a_- R\Sigma R^T\bv\\
    \bbeta' &=  R \bbeta + \bv - R\Sigma R^T \bv\\
    \Sigma' &= R \Sigma R^T.
  \end{split}
\end{equation}

Differentiating at the identity, we obtain the adjoint action of $G$
on $\g$.  We promote $(\alpha_+,\alpha_-,\balpha,\bbeta,\Sigma)$ to a
curve via the identity and simply compute the velocity at the identity
of the curve obtained after conjugation.  Doing so, we find that
\begin{equation}
  \Ad_{\sg(a_+,a_-,\ba,\bv,R)} \sA(x_+,x_-, \x,\y,X) = \sA(x'_+, x'_-, \x', \y', X'),
\end{equation}
where
\begin{equation}
  \label{eq:barg-adjoint}
  \begin{split}
   x'_+ &= x_+ - \tfrac12 x_-\|\bv\|^2 + \bv \cdot R \x - \ba \cdot  R\y + \bv \cdot R X R^T \ba\\
   x'_- &= x_-\\
   \x' &= R\x - x_- \bv - R X R^T (\ba + a_- \bv) + a_- R\y\\
   \y' &= R\y - RXR^T \bv\\
   X' &= RXR^T.
  \end{split}
\end{equation}

Let us introduce a basis $\lambda^{ab}, \beta^a, \pi^a,
\eta, \mu$ for $\g^*$ canonically dual to $L_{ab}, B_a, P_a, H, M$.
We now parametrise the dual $\g^*$ of the Bargmann Lie algebra
by\footnote{The choice of sign is such that later the momentum of a
  particle of mass $m$ moving with velocity $\bv$ is given by the
  familiar $\p = m \bv$.}
\begin{equation}
  \label{eq:bargmann-moment}
  \sM(m,E,\p,\bk,J) = m \mu + E \eta - p_a \pi^a + k_a \beta^a +
  \tfrac12 J_{ab} \lambda^{ab}.
\end{equation}
It follows that if $\alpha = \sM(m,E,\p,\bk,J)$, then the linear
function on $\g^*$ defined by $H$ takes the value $E$ on $\alpha$.
The dual pairing is given by
\begin{equation}
  \label{eq:barg-dual-pairing}
  \left<\sM(m,E,\p,\bk,J), \sA(x_+,x_-, \x, \y, X)\right>= m x_+ - E x_- - \p
  \cdot \x + \bk \cdot \y + \tfrac12 \Tr(J^TX).
\end{equation}
We then define the coadjoint representation as the dual representation
to the adjoint representation:
\begin{multline}
  \label{eq:coadjoint}
  \left<\Ad^*_{\sg(a_+,a_-,\ba,\bv,R)}\sM(m,E,\p,\bk,J), \sA(x_+,x_-, \x, \y,
    X)\right>\\ = \left<\sM(m,E,\p,\bk,J), \Ad_{\sg(a_+,a_-,\ba,\bv,R)^{-1}}\sA(x_+,x_-, \x, \y, X)\right>.
\end{multline}
We calculate
\begin{multline}
  \Ad_{\sg(a_+,a_-,\ba,\bv,R)^{-1}}\sA(x_+,x_-, \x, \y, X)\\
  = \Ad_{\sg(-a_+ + \ba \cdot \bv + \tfrac12 a_- \|\bv\|^2, -a_-, -R^T(\ba + a_-\bv), -R^T\bv, R^T)} \sA(x_+,x_-, \x,\y, X)
\end{multline}
to give
\begin{equation}
  \Ad_{\sg(a_+,a_-,\ba,\bv,R)^{-1}}\sA(x_+,x_-, \x, \y, X)  = \sA(x'_+, x'_-, \x', \y', X'),
\end{equation}
where
\begin{equation}
  \label{eq:adj-inv}
  \begin{split}
   x'_+ &= x_+ - \tfrac12 x_- \|\bv\|^2 - \bv \cdot \x + \ba \cdot \y + a_- \bv \cdot \y + \bv \cdot X\ba\\
   x'_- &= x_-\\
   \x' &= R^T(\x + X\ba + x_- \bv - a_- \y)\\
   \y' &= R^T(\y + X \bv)\\
   X' &= R^TXR.
  \end{split}
\end{equation}
The dual pairing then gives
\begin{equation}
  \Ad^*_{\sg(a_+,a_-,\ba,\bv,R)} \sM(m,E,\p,\bk,J) = \sM(m', E', \p', \bk', J'),
\end{equation}
where
\begin{equation}
  \label{eq:coajoint-action-expl}
  \begin{split}
    m' &= m \\
    E' &= E +  R\p \cdot \bv + \tfrac12 m \|\bv\|^2 \\
    \p' &= R\p + m \bv\\
    \bk' &= R \bk + m \ba + a_- (R\p + m\bv)\\
    J' &= RJR^T + \ba (R\p)^T - (R\p) \ba^T + (R\bk) \bv^T - \bv  (R\bk)^T - m \bv \ba^T + m \ba \bv^T.
  \end{split}
\end{equation}

\section{Maurer--Cartan one-form}
\label{sec:maurer-cartan-one}

In this section we let $g = \sg(a_+,a_-, \ba,\bv,R)$ be a generic
element of the Bargmann group and we will compute the pull-back of the
left-invariant Maurer--Cartan one-form to the parameter space:
\begin{equation}
  \begin{split}
      g^{-1} dg &=
  \begin{pmatrix}
    1 & -\tfrac12 \|\bv\|^2 & - \bv^T & - a_+ + \ba\cdot\bv + \tfrac12  a_- \|\bv\|^2\\
    0 & 1 & \bzero^T & -a_-\\
    0 & R^T\bv & R^T & - R^T(\ba + a_- \bv)\\
    0 & 0 & \bzero^T & 1
  \end{pmatrix}
  \begin{pmatrix}
    0 & - \bv\cdot d\bv & d\bv^T R + \bv^T dR & da_+\\
    0 & 0 & \bzero^T & da_-\\
    0 & -d\bv & dR & d\ba\\
    0 & 0 & \bzero^T & 0
  \end{pmatrix}\\
  &=
  \begin{pmatrix}
    0 & 0 & d\bv^T R & da_+ - \bv^T d\ba - \tfrac12 \|\bv\|^2 da_-\\
    0 & 0 & \bzero^T & da_-\\
    0 & -R^T d\bv & R^T dR & R^T\bv da_- + R^T d\ba\\
    0 & 0 & \bzero^T & 0
  \end{pmatrix}\\
  &= \sA(da_+- \bv^T d\ba - \tfrac12 \|\bv\|^2 da_-, da_-, R^T\bv da_- + R^T d\ba, R^Td\bv, R^TdR).
  \end{split}
\end{equation}
Pairing with $\alpha =\sM(m,E,\p,\bk,J)$, we find
\begin{equation}\label{eq:mc-form-alpha}
  \left<\alpha, g^{-1}dg\right> = m da_+ - (E + \tfrac12 m \|\bv\|^2 +  (R \p)^T \bv) da_- - (R\p + m \bv)^T d\ba + (R\bk)^T d\bv + \tfrac12 \Tr J^T R^TdR.
\end{equation}

\section{The case of $n=3$}
\label{sec:case-n=3}

When $n=3$, the vector and adjoint representations of $\SO(3)$ are
isomorphic.  As described in \cite{Figueroa-OFarrill:2023vbj}, the
isomorphism $\varepsilon: \RR^3 \to \so(3)$ is given by
$\varepsilon(\ba)\bb = \ba \times \bb$, which obeys
$[\varepsilon(\ba), \varepsilon(\bb)] = \varepsilon(\ba \times \bb)$
and also $\varepsilon(R\ba) = R \varepsilon(\ba) R^T$ and hence
$\ba \bb^T - \bb \ba^T = \varepsilon(\bb \times \ba)$.  It also
relates the inner products so that $\tfrac12 \Tr \varepsilon(\ba)^T
\varepsilon(\bb) = \ba \cdot \bb$.

It follows that in this dimension, letting $J = \varepsilon(\bj)$,
equation~\eqref{eq:coajoint-action-expl} becomes
\begin{equation}
  \label{eq:coadjoint-exp-3d}
  \begin{split}
    m' &= m \\
    E' &= E + R\p \cdot \bv + \tfrac12 m \|\bv\|^2 \\
    \p' &= R\p + m \bv\\
    \bk' &= R \bk + m \ba + a_- (R\p + m\bv)\\
    \bj' &=R\bj - \ba \times (R\p + m \bv) + \bv \times R\bk
  \end{split}
\end{equation}
and equation~\eqref{eq:mc-form-alpha} becomes
\begin{equation}
  \label{eq:mc-form-alpha-3d}
  \left<\alpha, g^{-1}dg\right> = m da_+ - (E +  R \p \cdot \bv + \tfrac12 m \|\bv\|^2) da_- - (R\p + m \bv) \cdot d\ba + R\bk\cdot d\bv +
  \bj \cdot \varepsilon^{-1}(R^{-1}dR).
\end{equation}

\section{Automorphisms}
\label{sec:automorphisms}

Let $A$ denote the group of automorphisms of the Bargmann Lie algebra
$\g$ which act trivially on the rotational subalgebra.  An easy
calculation shows that any $\varphi \in A$ is parametrised by
$\alpha,\beta \in \RR^\times$ and $\lambda,\mu \in \RR$ and acts on
the generators via
\begin{equation}
  \label{eq:autos}
  M \mapsto \alpha^2\beta M, \qquad H \mapsto \beta H + \lambda M,
  \qquad B_a \mapsto \alpha B_a + \mu P_a \qquad\text{and}\qquad P_a
  \mapsto \alpha\beta P_a.
\end{equation}
It is not hard to determine the dual action of $\varphi$ on $\g^*$:
$\varphi \cdot \sM(m,E,\p,\bk,\bj) = \sM(m',E',\p',\bk',\bj')$ where
\begin{equation}
  \label{eq:autos-on-coalgebra}
  \begin{split}
    m' &= \alpha^{-2}\beta^{-1} m \\
    E' &= \beta^{-1} E - \alpha^{-2}\beta^{-2} \lambda m \\
    \p' &= \alpha^{-1}\beta^{-1}\p \\
    \bk' &= \alpha^{-1}\bk + \alpha^{-2} \beta^{-1} \mu \p \\
    \bj' &=\bj.
  \end{split}
\end{equation}

The Lie algebra automorphism $\varphi$ integrates to an automorphism
$\Phi$ of the Bargmann Lie group $G$ via $\Phi(e^X) = e^{\varphi(X)}$
for $X \in \g$, which is well defined because the exponential map on
the nilpotent subgroup generated by $\bB,\bP,H,M$ is a
diffeomorphism.  A simple calculation yields
\begin{equation}
  \label{eq:G-auto}
  \Phi(\sg(a_+, a_-, \ba, \bv, R)) = \sg(\alpha^2\beta a_+ - \lambda
  a_- + \tfrac12 \alpha\mu v^2, \beta a_-, \alpha\beta \ba + \mu \bv,
  \alpha \bv, R).
\end{equation}

Unlike the case of the Carroll group treated in
\cite{Figueroa-OFarrill:2023vbj}, Bargmann automorphisms do not relate
different types of coadjoint orbits nor can they be used to relate
different classes of unitary irreducible representations.  We only
list them here for completeness.

\section{Coadjoint orbits for $n=3$}
\label{sec:coadjoint-orbits-3d}

From now on we will let $G$ denote the identity component of the
Bargmann group. As such, $G$ is diffeomorphic to
$\RR \times \RR \times \RR^3 \times \RR^3 \times \SO(3)$ and at the
group level all that happens is that the orthogonal transformation $R$
now has determinant $1$.

There are some obvious Casimirs of the Bargmann algebra. Clearly $M$
is one, which defines an invariant linear function on $\g^*$.  It is
therefore constant on coadjoint orbits, as we see from equation
\eqref{eq:coadjoint-exp-3d}.  There is also a quadratic Casimir
$\delta^{ab} P_a P_b - 2 H M$, which says that $\|\p\|^2 - 2 E m$ is a
constant on the orbits.  This provides a check of equation~\eqref{eq:coadjoint-exp-3d}:
\begin{equation}
  \begin{split}
    \|\p'\|^2 - 2 E' m' &= \|R\p + m\bv\|^2 - 2 (E + R\p \cdot \bv + \tfrac12 m \|\bv\|^2)m\\
    &= \|\p\|^2 + m^2 \|\bv\|^2 + 2 m R\p \cdot \bv - 2 E m - 2 m R\p \cdot \bv - m^2 \|\bv\|^2\\
    &= \|\p\|^2 - 2 E m.
  \end{split}
\end{equation}
Similarly, there is a quartic Casimir $\delta^{ab}W_a W_b$, where $W_a
= J_a M + \epsilon_{abc} P_b B_c$, which says that $\|m\bj + \p \times
\bk\|^2$ is constant on the orbits.  Again this provides another check of
equation~\eqref{eq:coadjoint-exp-3d}:
\begin{equation}
  \begin{split}
    m' \bj' - \p' \times \bk' &= m (R\bj - \ba \times (R\p + m \bv) + \bv \times R\bk) - (R\p + m \bv) \times (R \bk + m \ba + a_- (R\p + m\bv))\\
    &= m R \bj - m \ba \times (R\p + m \bv) + m \bv \times R\bk - R\p
    \times R\bk - m \bv \times R\bk - m  (R\p + m \bv) \times \ba\\
    &= R (m \bj - \p \times \bk),
  \end{split}
\end{equation}
where we have used that for $R \in \SO(3)$, $R\p \times R\bk = R(\p
\times \bk)$.  Since the vectors are related by a rotation, their
norms agree.

We separate orbits into two kinds depending on the value of the linear Casimir.

\subsection{Coadjoint orbits with $m\neq 0$}
\label{sec:coadj-orbits-with}

In this case, we may choose $\ba = - \tfrac1m R\bk $ and $\bv =
-\tfrac1m R\p$ in order to set $\p' = \bk' = \bzero$.  Doing so, we
bring $\sM(m,E,\p,\bk,\bj)$ to
\begin{equation}
  \sM(m, \tfrac1{2m} (\|\p\|^2 - 2 E m), \bzero, \bzero, \tfrac1m R(m \bj - \p \times \bk)),
\end{equation}
where we recognise the values of the quadratic Casimir and the vector
whose norm is the quartic Casimir.  For each pair $(E_0,w)$ consisting
of a real number $E_0$ and non-negative number $w$ we have a coadjoint
orbit with representative covector
\begin{equation}
  \sM(m, E_0 ,\bzero,\bzero,\frac{w}m \be_3) \in \g^*,
\end{equation}
where $(\be_1, \be_2, \be_3)$ is the standard orthonormal basis for
$\RR^3$, and where $E_0 = \tfrac1{2m}\|\p\|^2 - E$ and $w^2 =\|m \bj +
\p \times \bk\|^2$ are the values of the quadratic and quartic
Casimirs, respectively.

\subsection{Coadjoint orbits with $m = 0$}
\label{sec:coadj-orbits-m-zero}

In this case, the coadjoint action simplifies:
\begin{equation}
  \label{eq:coadjoint-exp-3d-m0}
  \begin{split}
    E' &= E + R\p \cdot \bv\\
    \p' &= R\p \\
    \bk' &= R (\bk + a_- \p) \\
    \bj' &=R\bj - \ba \times R\p + \bv \times R\bk.
  \end{split}
\end{equation}

The squared norm $\|\p\|^2$ is now invariant and we have two cases,
depending on whether it is zero or nonzero.

\subsubsection{The case $\p = \bzero$}
\label{sec:m-and-p-zero}

In this case both $E$ and $\|\bk\|^2$ are invariant and we must
distinguish between two cases:
\begin{enumerate}
\item If $\bk = 0$, then either $\bj = 0$ or else we may rotate it to
  any desired direction and we have orbits with representatives
  $\sM(0,E,\bzero,\bzero,j \be_3)$, where $j \geq 0$ and $E \in
  \RR$.
\item If $\bk \neq 0$, we may rotate it so that it points in the
  direction $\be_3$ we can then use $\bv$ to make $\bj'$ collinear to
  $\be_3$.  In other words, we have orbits with representatives
  $\sM(0,E,\bzero,k \be_3, j \be_3)$, where $k > 0$ and $E,j \in \RR$.
\end{enumerate}

\subsubsection{The case $\p \neq \bzero$}
\label{sec:m-zero-p-nonzero}

If $\p \neq \bzero$, we may use $\bv$ to set $E' =0$ and $a_-$ to make
$\bk'$ perpendicular to $\p$.  We may use $\ba$ in order to make $\bj'$
collinear with $\p$ but then we may use the component of $\bv$
perpendicular to $\p$ to bring $\bj'$ to $\bzero$.  Finally, using the
rotations which fix $\p$ we can bring $\bk$ to any desired direction
perpendicular to $\be_3$.  In other words, we have orbits with
representatives $\sM(0,0,p \be_3, k\be_2, \bzero)$, for $p>0$ and $k
\geq 0$.

\subsection{Summary}
\label{sec:summary}

We summarise the above discussion in Table~\ref{tab:coadjoint-orbits},
which lists the orbits together with the dimension and a set of
equations which determine the orbit as an algebraic submanifold of
$\g^*$.

\begin{table}[h]
  \centering
    \caption{Coadjoint orbits of the Bargmann group}
    \label{tab:coadjoint-orbits}
    \setlength{\extrarowheight}{3pt}
  \resizebox{\linewidth}{!}{
    \begin{tabular}{*{3}{>{$}l<{$}}>{$}c<{$}>{$}l<{$}}
      \multicolumn{1}{l}{\#} & \multicolumn{1}{c}{Orbit representative} & \multicolumn{1}{c}{Stabiliser} & \dim\mathcal{O}_\alpha & \multicolumn{1}{c}{Equations for orbits}\\
                             &
                               \multicolumn{1}{c}{$\alpha=\sM(m,E,\p,\bk,\bj)$} & \multicolumn{1}{c}{$G_\alpha$}& \\ \midrule \rowcolor{blue!7}
      1& \sM(m_0,E_0,\bzero,\bzero,\bzero) & \{\sg(a_+,a_-,\bzero,\bzero,R)\} & 6 &m=m_0\neq 0, \tfrac1{2m}(\|\p\|^2- 2 E m) = E_0, m\bj - \p\times\bk = \bzero\\
      2& \sM(m_0,E_0,\bzero,\bzero, s \be_3) & \{\sg(a_+,a_-,\bzero,\bzero,R) \mid R\be_3 = \be_3\} & 8 & m=m_0\neq 0, \tfrac1{2m}(\|\p\|^2- 2 E m) = E_0, \|m\bj - \p \times\bk \| = s >0\\\rowcolor{blue!7}
      3& \sM(0,E_0,\bzero,\bzero,\bzero) & G & 0 & m=0, E = E_0, \p=\bk = \bj = \bzero \\
      4& \sM(0,E_0,\bzero,\bzero,s\be_3) & \{\sg(a_+,a_-,\ba,\bv,R) \mid R\be_3 = \be_3\} & 2 & m=0, E = E_0, \p = \bk = \bzero, \|\bj\|=s>0 \\\rowcolor{blue!7}
      5& \sM(0,E_0,\bzero, k_0\be_3, h\be_3) & \{\sg(a_+,a_-,\ba,v \be_3,R) \mid R\be_3 = \be_3\} & 4 & m=0, E=E_0, \p = \bzero, \|\bk\|= k_0>0, \bj \cdot \bk = h k_0\\
      6& \sM(0,0,p_0\be_3,\bzero,\bzero) & \{\sg(a_+,0,a \be_3,\bv,R) \mid R\be_3 = \be_3, \bv \cdot \be_3 = 0 \} & 6 & m=0, \|\p\|=p_0>0, \p \times \bk = \bzero\\\rowcolor{blue!7}
      7& \sM(0,0,p_0\be_3, k_0\be_2,\bzero) & \{\sg(a_+,0,a \be_3, v \be_2,I)\} & 8 & m = 0, \|\p\| = p_0 >0, \|\p \times \bk\| = p_0 k_0>0\\
      \bottomrule
    \end{tabular}
  }
  \caption*{This table lists the different coadjoint orbits by
    exhibiting an orbit representative $\alpha\in \g^*$, its
    stabiliser subgroup $G_\alpha$ inside the Bargmann group, the
    dimension $\dim \eO_\alpha$ of the orbit and the equations
    defining the orbit.   In the last case, the second invariant is
    $\|\p \times \bk\|$, which takes the value $p_0 k_0$ only in the
    chosen representative.  In all cases one can check that $\dim
    \eO_\alpha = \dim G - \#\{\text{equations}\}$.}

\end{table}

\subsection{Coadjoint orbits of the full Bargmann group}
\label{sec:coadj-orbits-full-barg}

The full Bargmann group has two connected components since now $R \in
\Ort(3)$.  As a Lie group, $\Ort(3) = \SO(3) \cup \SO(3)P$, where the
parity $P$ can be thought of as space inversion, sending $\x \mapsto
-\x$ in $\RR^3$.  However it follows from the explicit form of the
orbit representatives, that their image under space inversion, which
sends $\sM(m,E,\p,\bk,\bj) \mapsto \sM(m,E,-\p,-\bk,\bj)$ lies in the
$\SO(3)$-orbit.  Therefore the coadjoint orbits in
Table~\ref{tab:coadjoint-orbits} are also the coadjoint orbits of the
full Bargmann group.  We may also extend the Bargmann group by time
reversal.  As shown, e.g., in
\cite[Section~II.D]{MR0466427}, time reversal acts on the Galilei
generators as $(H,P_a,B_a,L_a) \mapsto (-H,P_a,-B_a,L_a)$, which
extends by $M \mapsto -M$ to an automorphism of the Bargmann Lie
algebra.  This integrates to an automorphism of the Bargmann Lie group
and we can ask how it acts on its coadjoint orbits.  As shown, e.g.,
in \cite[App.~A.6]{Figueroa-OFarrill:2023vbj}, it maps coadjoint
orbits symplectomorphically into coadjoint orbits.  Under time
reversal, $\sM(m,E,\p,\bk,\bj) \mapsto \sM(-m,-E,\p,-\bk,\bj)$.  It is
then a simple matter to use the equations characterising the coadjoint
orbits in Table~\ref{tab:coadjoint-orbits} to see that time reversal
leaves invariant the orbits of types $\#6,7$ and pairs up two orbits
of the other types:
\begin{equation}
  \begin{split}
    &\sM(m_0,E_0,\bzero,\bzero,\bj)\quad\text{and}\quad \sM(-m_0,-E_0,\bzero,\bzero,\bj)\\
    &\sM(0,E_0,\bzero,\bzero,\bj)\quad\text{and}\quad \sM(0,-E_0,\bzero,\bzero,\bj)\\
    &\sM(0,E_0,\bzero,\bk,\bj)\quad\text{and}\quad \sM(0,-E_0,\bzero,\bk,-\bj).
  \end{split}
\end{equation}
It is natural to ask whether galilean field theories obey the CPT
theorem and the answer is negative \cite{MR1552511}.

\subsection{Structure of the orbits}
\label{sec:structure-orbits}

The Bargmann group is isomorphic to a semidirect product $(\SO(3)
\ltimes \RR^3) \ltimes \RR^5$, where the normal subgroup $\RR^5$ is
abelian and generated by $M,H,P_a$.  We can therefore use the results
in, e.g., Oblak's thesis \cite{Oblak:2016eij}, to describe the
coadjoint orbits geometrically.

As described in \cite{Figueroa-OFarrill:2023vbj}, the coadjoint orbit
$\eO_\alpha$ of $\alpha = (\kappa,\tau) \in \fk^* \oplus \ft^* = \g^*$
under $G = K \ltimes T$ with $T$ abelian are fibred products
\begin{equation}
  \begin{tikzcd}
    \eO_\alpha \arrow[r] \arrow[d] & T^*\eO_\tau \arrow[d] \\
    K\times_{K_\tau} \eO_{\kappa_\tau} \arrow[r] & \eO_\tau\\
  \end{tikzcd}
\end{equation}
where
\begin{itemize}
\item $\eO_\tau$ is the $K$-orbit of $\tau \in \ft^*$;
\item $K_\tau \subset K$ is the stabiliser of $\tau$, so that $\eO_\tau \cong K/K_\tau$;
\item $\eO_{\kappa_\tau}$ is the $K_\tau$-coadjoint orbit of the
  restriction $\kappa_\tau \in \fk_\tau^*$ of $\kappa \in \fk^*$ to
  the Lie algebra $\fk_\tau$ of $K_\tau$.
\end{itemize}
The standard notation for such fibred products is
\begin{equation}
  \label{eq:fibred-product}
  \eO_\alpha = T^*\eO_\tau \times_{\eO_\tau} \left( K\times_{K\tau} \eO_{\kappa_\tau} \right),
\end{equation}
whose dimension can be read off as follows:
\begin{equation}
  \dim\eO_\alpha = \dim T^*\eO_\tau - \dim \eO_\tau + \dim K - \dim K_\tau + \dim \eO_{\kappa_\tau} = 2 \dim \eO_\tau + \dim \eO_{\kappa_\tau},
\end{equation}
which is of course even dimensional since $\eO_{\kappa_\tau}$ is
itself a coadjoint orbit of $K_\tau$.  In
Table~\ref{tab:structure-orbits} we deconstruct the coadjoint orbits
in Table~\ref{tab:coadjoint-orbits}.

\begin{table}[h]
  \centering
    \caption{Deconstructing the coadjoint orbits}
    \label{tab:structure-orbits}
  \begin{adjustbox}{max width=\textwidth}
    \begin{tabular}{l|*{8}{>{$}c<{$}}}
      \# & \alpha \in \g^* & \tau \in \ft^* & \eO_\tau & K_\tau & \kappa \in \fk^* & \kappa_\tau \in \fk_\tau^* & \eO_{\kappa_\tau} & \eO_\alpha\\
      \toprule\rowcolor{blue!7}
      1& \sM(m,E,\bzero,\bzero,\bzero)_{m\neq0,E\in\RR} & (m,E,\bzero) & \RR^3 & \SO(3) & (\bzero,\bzero) & \bzero &  \{\bzero\} & T^*\RR^3 \\
      2& \sM(m,E,\bzero,\bzero,\bj)_{m\neq0,E\in\RR,\bj\neq \bzero} & (m,E,\bzero) & \RR^3 & \SO(3) & (\bzero,\bj) & \bj &  S^2_{\|j\|} & T^*\RR^3 \times_{\RR^3} (K \times_{K_\tau} S^2) \\\rowcolor{blue!7}
      3& \sM(0,E,\bzero,\bzero,\bzero) & (0,E,\bzero) & \{(0,E,\bzero)\} & K & (\bzero,\bzero) & (\bzero,\bzero) &  \{(\bzero,\bzero)\} & \{(0,E,\bzero,\bzero,\bzero)\}\\
      4& \sM(0,E,\bzero,\bzero,\bj)_{E\in\RR,\bj\neq\bzero} & (0,E,\bzero) & \{(0,E,\bzero)\} & K & (\bzero,\bj) & (\bzero,\bj) & S^2_{\|\bj\|}  & S^2 \\\rowcolor{blue!7}
      5& \sM(0,E,\bzero,\bk,\bj)_{E\in\RR,\bk\times\bj=\bzero,\bk\neq\bzero} & (0,E,\bzero) & \{(0,E,\bzero)\} & K & (\bk,\bj) & (\bk,\bj) & T^*S^2_{\|\bk\|} & T^*S^2\\
      6& \sM(0,0,\p,\bzero,\bzero)_{\p\neq \bzero} & (0,0,\p) & \RR \times S^2_{\|\p\|} & \RR^2 \rtimes \SO(2) & (\bzero,\bzero) & (\bzero,\bzero) & \{(\bzero,\bzero)\} & T^*(\RR \times S^2)\\\rowcolor{blue!7}
      7& \sM(0,0,\p,\bk,\bzero)_{\bk \cdot \p =0, \bk \times \p \neq
         \bzero} & (0,0,\p) & \RR \times S^2_{\|\p\|} & \RR^2 \rtimes \SO(2) & (\bk,\bzero) & (\bk^\perp,\bzero) & T^*S^1_{\|\bk^\perp\|} & T^*(\RR \times S^2) \times_{\RR \times S^2} (K \times_{K_\tau} T^*S^1)\\
      \bottomrule
    \end{tabular}
  \end{adjustbox}
  \vspace{1em}
  \caption*{The manifold $\RR^3$ in cases \#1,2 is embedded in
    $\ft^*\cong \RR^5$ as the paraboloid
    $\left\{\sM(m, E + \frac1{2m}\|\p\|^2,\p) \mid \p \in
      \RR^3\right\}$. The stabiliser $K_\tau$ in cases \#6,7 consists
    of elements $\sg(\bv,R) \in K$ where $\bv \perp \p$ and $R\p = \p$.
    Since $\p \neq \bzero$, these are rotations about the axis defined
    by $\p$ and hence isomorphic to $\SO(2)$, so that the stabiliser
    is isomorphic to $\RR^2 \rtimes \SO(2)$.  The $K_\tau$-coadjoint
    orbits should be self-explanatory.  In cases \#3,4,5, the
    stabiliser is $K \cong \ISO(3)$, whose coadjoint orbits can be
    read off from \cite[App.~B.5.2]{Figueroa-OFarrill:2023vbj}: in case
    \#3 we have the point-like orbit $\{(\bzero,\bzero)\}$, in case \#4
    we have the 2-sphere of radius $\|\bj\|$, and in case \#5 we have
    the cotangent bundle of the sphere of radius $\|\bk\|$.  In cases
    \#6,7, the stabiliser $K_\tau$ is isomorphic to $\ISO(2)$ and the
    coadjoint orbits again can be read off from
    \cite[App.~B.5.2]{Figueroa-OFarrill:2023vbj}.  In case \#6 we have
    the point-like orbit $\{(\bzero,\bzero)\}$ and in case \#7 we have
    the cylinder $T^*S^1_{\|\bk^\perp\|}$, where $\bk^\perp$ is the
    component of $\bk$ perpendicular to $\p$ -- that being the
    restriction of $\kappa$ to $\fk_\tau$.}
\end{table}

\section{Actions of Galilei particles}
\label{sec:action-galil}

In this section we discuss the actions for Galilei particles. As
expected, we will recover the well-known particle action for massive
Galilei particles, but we will also cover the possible less familiar
massless case in full generality. These actions provide information
concerning the mobility of the particles and are the starting point
for many applications, e.g., for path integral
quantisation~\cite{Alekseev:1988vx}.

As discussed above, classical Galilei particles correspond to
homogeneous symplectic manifolds of the Galilei group and these in
turn correspond to coadjoint orbits of the Bargmann group. Hence they
can also be thought of as classical Bargmann particles. We view the
Bargmann group as an auxiliary concept we are forced to introduce for
mathematical consistency, but from a spacetime perspective it is the
Galilei group which is the relativity group and hence we prefer to use
the term Galilei particle, but in so doing we allow them to have a
non-zero mass.

Let $\alpha \in \g^*$ be an element in the dual of the Bargmann
algebra and let $\eO_\alpha \subset \g^*$ denote its coadjoint orbit.
It is $G$-equivariantly diffeomorphic to $G/G_\alpha$, with $G_\alpha$
the stabiliser subgroup of $\alpha$.  Let $\pi_\alpha : G \to
\eO_\alpha$ denote the orbit map: $\pi_\alpha(g)= \Ad^*_g \alpha$.
The $G$-invariant Kirillov--Kostant--Souriau symplectic structure
$\omega_{\text{KKS}}$ on $\eO_\alpha$ pulls back via the orbit map to
a left-invariant presymplectic form $\pi_\alpha^*\omega_{\text{KKS}}
\in \Omega^2(G)$ on $G$, which is moreover exact:
\begin{equation}
  \pi_\alpha^*\omega_{\text{KKS}} = - d \left<\alpha, \vartheta\right>,
\end{equation}
where $\vartheta \in \Omega^1(G;\g)$ is the left-invariant
Maurer--Cartan one-form on $G$. The primitive one-form
$\left<\alpha, \vartheta\right>$ defines a variational problem for
curves $g: I \to G$ in the group:
\begin{equation}
  \label{eq:action-for-alpha}
  S[g]:= \int_I \left<\alpha, g^*\vartheta\right> = \int_I
  \left<\alpha, g(\tau)^{-1}\dot g(\tau)\right> d\tau,
\end{equation}
where $\tau \in I$ is the parameter along the curve and should not be confused with the element of $\ft^*$ used in the previous section. This is the point of departure in this section for the study of
the dynamical systems associated to each of the coadjoint orbits.
Some of these actions have been discussed, e.g., in
\cite[§5.2]{Bergshoeff:2022eog}.

We will now construct actions $S=\int L d\tau$ for the Galilei
particles in $3+1$ dimensions. Using~\eqref{eq:mc-form-alpha-3d} with
the replacements $a_{-} \mapsto t$ and $\ba \mapsto -\x$ leads to the
general Lagrangian
\begin{subequations}
    \label{eq:mc-form-alpha-action}
  \begin{align}
    L[a_{+},t,\x,\bv,R(\bm{\varphi})]&= m (\dot a_+ +  \bv \cdot \dot {\x} - \tfrac12  \|\bv\|^2 \dot t) - E \dot t + R\p \cdot ( \dot {\x}- \bv \dot t)+ R\bk\cdot \dot \bv  +
                                       \bj \cdot \varepsilon^{-1}(R^{-1}\dot R)\\
                                     &= m \dot a_+ - (E +  R \p \cdot \bv + \tfrac{m}{2} \|\bv\|^2) \dot t + (R\p + m \bv) \cdot \dot {\x} + R\bk\cdot \dot \bv +
                                       \bj \cdot \varepsilon^{-1}(R^{-1}\dot R) 
  \end{align}
\end{subequations}
where the dot denotes derivatives with respect to the parameter
$\tau$. We parametrise the orbits by $\alpha =\sM(m,E,\p,\bk,J)$ which
means they are not varied, while we vary with respect to the
quantities in the square brackets of the Lagrangian. Since the action
does not depend on the specific point $\alpha$, but only on the
coadjoint orbit itself, we are free to make a convenient choice. We
will use the representatives in Table~\ref{tab:coadjoint-orbits}.
Since the physics and degrees of freedom depend on the specific
particle we will analyse them case by case, but first we discuss the
global and gauge symmetries for the generic
Lagrangian~\eqref{eq:mc-form-alpha-action}.

\subsection{Symmetries}
\label{sec:symmetries}

The action~\eqref{eq:mc-form-alpha-action} has global Galilei symmetry
since $g^{-1}\dot g$ is invariant under the $\tau$-independent left
action $g \mapsto h g$. Consequentially, the infinitesimal symmetries
lead to Noether charges $\frac{d}{d\tau} X_{Q} = 0$ which are given by
\begin{subequations}
  \label{eq:symmetries-general}
  \begin{align}
    \delta_{c_{+}} a_{+} &= c_{+}  &&&&&    m_{Q} &= m \\
    \delta_{c_{t}} t &= c_{t}  &&&&&    E_{Q} &= E + R\p \cdot \bv + \tfrac12 m \|\bv\|^2 \\
    \delta_{c_{x}} \x &= \bm{c}_{x} &&&&& \Rightarrow\quad   \p_{Q} &= R\p + m \bv\\
    \delta_{c_{v}} \bv &=\bm{c}_{v} &\delta_{c_{v}} a_{+} &= -\x \cdot \bm{c}_{v} & \delta_{c_{v}} \x &=t \bm{c}_{v} &    \bk_{Q} &= R \bk - m \x + t (R\p + m\bv)\\
    \delta_{\omega} R &= \omega R & \delta_{\omega} \x &= \omega \x & \delta_{\omega} \bv &= \omega \bv  &  \bj_{Q} &=R\bj +\x \times (R\p + m \bv) + \bv \times R\bk
  \end{align}
\end{subequations}
where $\omega^{T} = -\omega$. This shows that the charges are given by
the coadjoint action on $\alpha$, cf.,~\eqref{eq:coadjoint-exp-3d}.

This action also has gauge freedom parametrised by the right action
$g \mapsto g h(\tau)$, where $h$ is now $\tau$ dependent and has to be
in the stabiliser of $\alpha$. The general infinitesimal gauge
transformations for the case at hand are given by
\begin{subequations}
  \label{eq:gaugeinv-general}
  \begin{align}
    \delta_{\lambda_{+}} a_{+}  & = \lambda_{+} \\
    \delta_{\lambda_{t}} t      & = \lambda_{t} & \delta_{\lambda_{t}}a_{+}  & = - \tfrac12 \|\bv\|^2 \lambda_{t} & \delta_{\lambda_{t}}\x & =\bv \lambda_{t}                \\
    \delta_{\lambda_{x}} \x     & = R \bm{\lambda}_{x}  & \delta_{\lambda_{x}} a_{+}         & = - \bv \cdot R \bm{\lambda}_{x}           \\
    \delta_{\lambda_{v}} \bv    & = R \bm{\lambda}_{v}  \\
    \delta_{\lambda_{\omega}} R & = R\lambda_{\omega}                                                                                            
  \end{align}
\end{subequations}
where all $\lambda$ are $\tau$-dependent and
$\lambda_{\omega}^{T} = -\lambda_{\omega}$.

\subsection{Massive Galilei particles}
\label{sec:mass-galil-part}

In this subsection, we construct Lagrangians associated with massive
orbits, both without spin (orbit $\#1$) and with spin (orbit $\#2$).

\subsubsection{Orbit $\#1$ (massive spinless)}
\label{sec:orbit-massive-no-spin}

The massive orbit without spin describes the most familiar type of
galilean particle commonly encountered in textbooks. We will use this
section to illustrate some known properties of these geometric actions
(see,
e.g.,~\cite{Oblak:2016eij,Barnich:2022bni,Figueroa-OFarrill:2023vbj,Basile:2023vyg}
and references therein) and provide further details in
Appendix~\ref{sec:symm-mass-spinl}.

According to Table~\ref{tab:coadjoint-orbits}, massive spinless
Galilei particles can be characterised by the following representative
\begin{equation}
  \label{eq:rep_orbit1}
  \alpha=\sM(m,E_{0},\boldsymbol{0},\boldsymbol{0},\boldsymbol{0}) \, .
\end{equation}
Using this representative in~\eqref{eq:mc-form-alpha-action} leads to
the following action for the massive galilean particle
\begin{equation}
  \label{eq:Lagr-start}
 L[a_{+},t,\x,\bv,R(\bm{\varphi})] = m \dot a_+ - (E_{0} + \tfrac12 m
  \|\bv\|^2) \dot t  + m \bv \cdot \dot \x \,.
\end{equation}
The first term in~\eqref{eq:Lagr-start} is a boundary term that
ensures the existence of a non-vanishing conserved quantity $m$. This
term and the variation with respect to $R(\bm{\varphi})$ do not
contribute to the equations of motion and we will therefore omit it in
the following and concentrate our discussion on the following
Lagrangian
\begin{equation}
  \label{eq:Lagr}
L[t,\x,\bv] =  - (E_{0} + \tfrac12 m \|\bv\|^2) \dot t  + m \bv \cdot \dot \x \,.
\end{equation}

To express the Lagrangian in canonical form we introduce the canonical
momenta
\begin{subequations}
\begin{align}
  \label{eq:momenta}
  p_{t} &= \frac{\partial L}{\partial \dot{t}} = - (E_{0} + \tfrac12 m \|\bv\|^2)\\
  \bm{p}&=\frac{\partial L}{\partial \dot{\x}}= m \bv \\
  \bm{p}_{v}&=\frac{\partial L}{\partial \dot{\bv}}=\bzero \,,
\end{align}
\end{subequations}
which lead to the following constraints
\begin{subequations}
\begin{align}
  \phi &= p_{t} +E_{0} + \tfrac12 m \|\bv\|^2\approx 0 \\
  \bm{\phi}^{1} &=\bm{p} - m\bv\approx \boldsymbol{0} \\
  \bm{\phi}^{2} &= \bm{p}_{v} \approx \boldsymbol{0} \,  .
\end{align}
\end{subequations}
We can use them to construct an action in Hamiltonian form, which is
generically of the form
\begin{align}
  \int L_{\mathrm{can}}[q,p,u,\phi] d\tau =\int (p\dot q - H_{\text{can}} - u \phi) d\tau  
\end{align}
with Poisson brackets given by $\{q,p\}=1$. For the case at hand this
leads to
\begin{align}
  \label{eq:massive-canonical-full}
  L_{\mathrm{can}}[t,p_{t},\x,\bm{p},\bv,\bm{p}_{v},u, \bm{u}_{1},\bm{u_{2}}]
  = p_{t}\dot t + \bm{p}\cdot \dot \x + \bm{p}_{v} \cdot \dot{\bm{v}} - u \phi - \bm{u}_{i} \bm{\phi}^{i} \, ,
\end{align}
where we observe the vanishing of the canonical Hamiltonian
$H_{\text{can}}$ and the enforcement of constraints through the
variation of the Lagrange multipliers $u$ and $\bm{u}_{i}$. 
The set of constraints $\left(\bm{\phi}^{1},\bm{\phi}^{2}\right)$ are
of second-class. Indeed, they obey the following non-vanishing Poisson
brackets
\begin{equation}
    \left\{ \phi^{1}_{a},\phi^{2}_{b}\right\} = - m\delta_{ab} \,.
\end{equation}
The second class constraints can be imposed to be strongly equal to
zero. In particular, the constraint $\bm{\phi}^{1}=0$ can be
conveniently solved as $\bv = \bm{p}/m$.  Thus, plugging it back in the action one obtains
\begin{equation}
  \label{eq:reparem}
  L_{\mathrm{can}}[t,p_{t},\x,\bm{p},u]
  = p_{t}\dot t + \bm{p}\cdot \dot \x - u 
  \left(
   p_{t} + \frac{1}{2m} \| \bm{p}\|^2 + E_{0} 
  \right) .
\end{equation}
We could have circumvented the analysis of the second class
constraints by realising that this part of the
action~\eqref{eq:Lagr-start} is already in first order form, i.e., we
could have just redefined $\bv = \bm{p}/m$ in~\eqref{eq:Lagr-start}.

On the other hand, the constraint $\phi$ is of first-class and
generates the gauge symmetry of time reparametrisations. By solving
this first-class constraint and applying the gauge-fixing condition
$t=\tau$, the action can be written as
\begin{align}
  \label{eq:massive-can-final}
  L_{\mathrm{can}}[\x,\bm{p}]=\bm{p}\cdot \dot \x  -
  \left(
  \frac{1}{2m} \| \bm{p}\|^2 +E_{0}  
  \right) ,
\end{align}
where the derivatives are now with respect to $t$. To write this
action in configuration space we can use the equation of motion
obtained from the variation of $\bm{p}$ to obtain
\begin{align}
  \label{eq:massive-conf}
  L[\x]_{\mathrm{red}}=  \frac{m}{2} \| \dot \x\|^2 - E_{0} \,.
\end{align}
This is the standard action for a free (nonrelativistic) Galilei
particle with mass $m$.

The dimension of the orbit \#2 in Table~\ref{tab:coadjoint-orbits}
indeed agrees with the number of independent canonical variables of
our actions. From~\eqref{eq:massive-canonical-full} we obtain that
they are $14-2\times 1 - 6 = 6$, where we have taken all canonical
variables ($14$) and subtracted the constraints (first-class
constraints count twice, e.g., Section 1.4.2.\
in~\cite{Henneaux:1992ig}). This also agrees with the $6$ canonical
variables in~\eqref{eq:massive-can-final}, where all constraints have
been resolved.

\subsubsection{Orbit $\#2$ (massive spinning)}
\label{sec:orbit-massive-spin}

The representative of this orbit is given by
\begin{equation} 
\alpha=\sM(m,E_{0},\boldsymbol{0},\boldsymbol{0},\boldsymbol{j}).\label{eq:rep_orbit2}
\end{equation}
From~\eqref{eq:mc-form-alpha-action} one obtains the following
Lagrangian
\begin{equation}
  \label{eq:Lagr2}
L[t,\x,\bv,R\left(\boldsymbol{\varphi}\right)] =  - \left(E_{0} + \tfrac12 m \|\bv\|^2\right) \dot t  + m \bv \cdot \dot \x +
  \bj \cdot \varepsilon^{-1}(R^{-1}\dot{R}) \,.
\end{equation}
An important property is that the last term, that describes the spin part of the particle, ``decouples''
from the rest of the action. This can be seen as a
consequence of the fact that there is no Thomas precession for
galilean particles (the same is true for carrollian
particles~\cite{Figueroa-OFarrill:2023vbj}). On the other hand, the first two terms at the right-hand side are identical to those discussed in orbit $\#1$. 

Following~\cite{Figueroa-OFarrill:2023vbj}, it is convenient to
parametrise the rotation matrix as follows
\begin{equation}
R\left(\boldsymbol{\varphi}\right)=e^{\varphi_{1}\epsilon_{1}}e^{\varphi_{2}\epsilon_{2}}e^{\varphi_{3}\epsilon_{3}},\label{eq:Rot}
\end{equation}
where $(\epsilon_{a})_{bc}=-\epsilon_{abc}$. Furthermore, if we choose the angular momentum to be aligned with the $z$-axis, i.e. $\boldsymbol{j}=\left(0,0,s\right)$ with $s>0$, then the Lagrangian can be written as 
\begin{equation}
  \label{eq:Lagr3}
L[t,\x,\bv,\boldsymbol{\varphi}] =  - \left(E_{0} + \tfrac12 m \|\bv\|^2\right) \dot t  + m \bv \cdot \dot \x +s\left(\dot{\varphi}_{3}+\sin\left(\varphi_{2}\right)\dot{\varphi}_{1}\right).
\end{equation}
To express the Lagrangian in canonical form, we can use the momenta
defined in~\eqref{eq:momenta} and the momenta associated with the
spin part, which are given by
\begin{equation}
\Pi_{1}=\frac{\partial L}{\partial\dot{\varphi}_{1}}=s\sin\varphi_{2} \,,\quad \Pi_{2}=\frac{\partial L}{\partial\dot{\varphi}_{2}}=0 \,,\quad \Pi_{3}=\frac{\partial L}{\partial\dot{\varphi}_{3}}=s \,,
\end{equation}
Following the same approach as for orbit $\#1$ for the non-spinning part, the Lagrangian in canonical form can be written as 
\begin{align}
  \label{eq:reparamspinn}
  L_\text{can}[t,p_{t},\x,\bm{p},\boldsymbol{\varphi},\boldsymbol{\Pi},u,u_2,u_3]
  = p_{t}\dot t + \bm{p}\cdot \dot \x +\boldsymbol{\Pi}\cdot\dot{\boldsymbol{\varphi}}- u 
  \left(
   p_{t}+ \frac{1}{2m} \| \bm{p}\|^2  +E_{0} 
  \right) -u_2\Pi_2-u_3\left(\Pi_3-s\right).
\end{align}
To emphasise the relevant physical degrees of freedom, we can solve the constraint and impose the gauge fixing condition $t=\tau$. Then, after eliminating some boundary terms, the Lagrangian in the reduced phase space takes the form 
\begin{align}
  \label{eq:massive-spin-can}
  L_{\text{red}}[\x,\bm{p},\varphi_{1},\Pi_{1}]=\bm{p}\cdot \dot \x  +\Pi_{1} \dot{\varphi}_{1}-
  \left(
 \frac{1}{2m} \| \bm{p}\|^2 + E_{0} 
  \right) .
\end{align}
Alternatively, by eliminating the linear momentum using its equation of motion, we can express the Lagrangian for the spinning massive Galilei particle as follows:
\begin{align}
  \label{eq:masive-spin-conf}
  L[\x,\varphi_{1},\Pi_{1}]=  \frac{m}{2} \| \dot \x\|^2+\Pi_{1} \dot{\varphi}_{1} -  E_{0}  \, .
\end{align}

Comparing our actions~\eqref{eq:massive-spin-can} or
\eqref{eq:masive-spin-conf} with Section~\ref{sec:mass-galil-part} we
see that they have two additional canonical variables and hence in
total $8$ independent canonical variables, which agrees with
Table~\ref{tab:coadjoint-orbits}.

\subsection{Massless Galilei particles}
\label{sec:massl-galil-part}

In this subsection, we construct the Lagrangians and study dynamics of
massless Galilei particles. The foundational aspects for the analysis
were provided in~\cite{MR0260238, MR1066693}, and the dynamics
associated with the orbit \#6 was discussed in~\cite{MR0260238,
  Bergshoeff:2022eog, Bergshoeff:2022qkx}. Here, we present a
self-contained discussion of this case, while also extending our
analysis to include those orbits that have not been explored
previously.

From a galilean perspective, although massless Galilei particles may
not seem to describe any known particle, they are however connected to
geometrical optics~\cite{Duval:2005ry} and they emerge as the most
relevant entities in the context of their application to planons. In
this scenario, they represent elementary dipoles with restricted
motion. A comprehensive discussion of this approach will be presented
in our forthcoming work~\cite{Planons:2024}.

\subsubsection{Orbit $\#3$ (vacuum)} 
According to Table~\ref{tab:coadjoint-orbits} the representative of this orbit is given by 
\begin{equation} 
\alpha=\sM(0,E_{0},\boldsymbol{0},\boldsymbol{0},\boldsymbol{0}) \,. \label{eq:rep_orbit3}
\end{equation}
The corresponding Lagrangian can then be directly obtained using the
orbit representative in~\eqref{eq:mc-form-alpha-action}. It is given
by
\begin{equation*}
L[t]=-E_{0}\dot{t} \,,
\end{equation*}
and is a pure boundary term. Considering the trivial dynamics and the fact that the stabiliser is the entire Bargmann group, one can interpret this orbit as the vacuum configuration.

\subsubsection{Orbit $\#4$ (spinning vacuum)}

The orbit representative for this case is given by
\begin{equation}
\alpha=\sM(0,E_{0},\boldsymbol{0},\boldsymbol{0},\boldsymbol{j}) \,. \label{eq:rep_orbit4}
\end{equation}
Thus, using~\eqref{eq:mc-form-alpha-action} one finds
\begin{equation}
L\left[t,R\left(\boldsymbol{\varphi}\right)\right]=-E_{0}\dot{t}+ \bj \cdot \varepsilon^{-1}(R^{-1}\dot{R}) \,.
\end{equation}
The first term at the right-hand side is a boundary term that can be
neglected, while the second one describes the spin degrees of freedom.
Thus, if the angular momentum is aligned with the $z$-axis, i.e.,
$\boldsymbol{j}=\left(0,0,s\right)$, and if one employs the same
parameterisation for the rotations as introduced in
Eq.~\eqref{eq:Rot}, then the Lagrangian becomes
\begin{equation}
L\left[\boldsymbol{\varphi}\right]=s\left(\dot{\varphi}_{3}+\sin\left(\varphi_{2}\right)\dot{\varphi}_{1}\right) .
\end{equation}
Therefore, this configuration may be interpreted as a spinning vacuum.

\subsubsection{Orbit $\#5$}
\label{sec:Lag_orbit_5}

This orbit is determined by the following representative:
\begin{equation} 
\alpha=\sM(0,E_{0},\boldsymbol{0},\boldsymbol{k},\boldsymbol{j}) \,, \label{eq:rep_orbit5}
\end{equation}
where $\left\Vert \boldsymbol{k}\right\Vert =k_{0}>0$ and
$\boldsymbol{j}\cdot\boldsymbol{k}=hk_{0}$.
Using~\eqref{eq:mc-form-alpha-action} one finds the following
Lagrangian
\begin{equation}
  L\left[t,\boldsymbol{v},R\left(\boldsymbol{\varphi}\right)\right]=-E_{0}\dot{t}+R\boldsymbol{k}\cdot\dot{\boldsymbol{v}}+\bj \cdot \varepsilon^{-1}(R^{-1}\dot{R}) \,.
\end{equation}
It is convenient to consider the parametrisation in Eq.~\eqref{eq:Rot}
for the rotation matrix, and to write
$\boldsymbol{k}=k_{0}\hat{\boldsymbol{n}}$, where
$\hat{\boldsymbol{n}}$ is the unit vector defined by
\begin{equation}
\hat{\boldsymbol{n}}=\left(\sin\varphi_{2},-\sin\varphi_{1}\cos\varphi_{2},-\cos\varphi_{1}\cos\varphi_{2}\right) .
\end{equation}
Therefore, when the angular momentum aligns with the $z$-axis
($\boldsymbol{j}=\left(0,0,h\right)$), the Lagrangian becomes
\begin{equation}
  L\left[\boldsymbol{\varphi},\boldsymbol{v},t\right]=-E_{0}\dot{t}+k_{0}\hat{\boldsymbol{n}}\cdot\dot{\boldsymbol{v}}+h\left(\dot{\varphi}_{3}+\sin\left(\varphi_{2}\right)\dot{\varphi}_{1}\right).
\end{equation}
Indeed, up to boundary terms and the renaming
$\boldsymbol{k} \rightarrow \boldsymbol{p}$ and
$\boldsymbol{v}\rightarrow \boldsymbol{x}$, this action is identical
to the one found in the study of coadjoint orbits of the Carroll
group, referred to as ``massless Carrollion'' in Ref.
\cite{Figueroa-OFarrill:2023vbj}. The reason is that the space
defining the orbits are the same in both cases.

The canonical form of the action is obtained by introducing the
canonical momenta
\begin{equation}
  p_{t}=\frac{\partial L}{\partial\dot{t}}=-E_{0} \,,\quad\boldsymbol{p_{v}}=\frac{\partial L}{\partial\dot{\boldsymbol{v}}}=k_{0}\hat{\boldsymbol{n}} \,,\quad \Pi_{1}=\frac{\partial L}{\partial\dot{\varphi}_{1}}=h\sin\varphi_{2} \,,\quad \Pi_{2}=\frac{\partial L}{\partial\dot{\varphi}_{2}}=0 \,,\quad \Pi_{3}=\frac{\partial L}{\partial\dot{\varphi}_{3}}=h \,,
\end{equation}
which satisfy the following constraints
\begin{equation}
\left\Vert \boldsymbol{p_{v}}\right\Vert ^{2}-k_{0}^{2}=0 \,,\quad k_{0}\Pi_{1}-h p_{v}^{1}=0 \,,\quad \Pi_{2}=0 \,,\quad \Pi_{3}-h=0 \,.
\end{equation}
These constraints are of first class. Therefore, neglecting boundary
terms, the Lagrangian in canonical form can be written as
\begin{align}
  L_{\text{can}}\left[\boldsymbol{\varphi},\boldsymbol{\Pi},\boldsymbol{v},\boldsymbol{p_{v}},t,E,u,u^{1},\boldsymbol{\eta}\right] & =-E\dot{t}+\boldsymbol{\Pi}\cdot\dot{\boldsymbol{\varphi}}+\boldsymbol{p_{v}}\cdot\dot{\boldsymbol{v}}-u\left(E-E_{0}\right)-u_{1}\left(\left\Vert \boldsymbol{p}_{v}\right\Vert^{2}-k_{0}^{2}\right) \nonumber\\
&\quad -\eta_{1}\left(k_{0}\Pi_{1}-hp_{v}^{1}\right)-\eta_{2}\Pi_{2}-\eta_{3}\left(\Pi_{3}-h\right).
\end{align}
For simplicity, let us restrict to the case with vanishing spin
($h=0$) and let us fix the gauge $t=\tau$. Then, the Lagrangian
becomes
\begin{equation}
L_{\text{can}}\left[\boldsymbol{v},\boldsymbol{p_{v}},u_{1}\right]=\boldsymbol{p_{v}}\cdot\dot{\boldsymbol{v}}-u_{1}\left(\left\Vert \boldsymbol{p_{v}}\right\Vert ^{2}-k_{0}^{2}\right)-E_{0} \,,
\end{equation}
where a dot now stands for derivative with respect to the physical
time $t$.

Next, we can solve for $\p_v$. By varying with respect to $\p_v$ and
$u_1$, we find
\begin{equation}
    \Vert\p_v\Vert^2 = k_0^2 \,,\quad 2 u_1 \p_v = \dot \bv \,.
\end{equation}
These equations are solved by writing $\p_v = k_0 \hat{\bm{n}}$, and
$u_1 = \frac{1}{2k_0} \Vert\bv\Vert$. Plugging back in the Lagrangian
we obtain
\begin{equation}
    L_\text{red}[\bv] = -E_0 + k_0\Vert\dot \bv\Vert \,.
\end{equation}
As a final remark, the counting of independent variables from the Hamiltonian analysis $14 - 2 \times 5 = 4$ coincides with the dimension of this orbit.

\subsubsection{Orbit $\#6$} 

The representative of this orbit is given by
\begin{equation} 
\alpha=\sM(0,0,\boldsymbol{p},\boldsymbol{0},\boldsymbol{0}),\label{eq:rep_orbit6}
\end{equation}
where $\Vert\p\Vert = p_0 > 0$. From~\eqref{eq:mc-form-alpha-action}
one finds the following Lagrangian:
\begin{equation}
L\left[\boldsymbol{\varphi},\boldsymbol{v},\boldsymbol{x},\mathit{t}\right]=\left(R\boldsymbol{p}\cdot\boldsymbol{v}\right)\dot{t}+\left(R\boldsymbol{p}\right)\cdot\dot{\boldsymbol{x}} \,. \label{eq:Lagrangian_orbit_6}
\end{equation}
It can be written in canonical form as follows
\begin{equation}
    L_\text{can}[t,p_t,\x,\bpi,\bv,\p_v,u,u_1,\bm{u}_2] = p_t \dot t + \bpi \cdot \dot \x + \p_v \cdot \dot \bv - u \phi - u_1 \phi^1 - \bm{u}_2 \cdot \bm{\phi}^2 \,,\label{eq:Lcan6}
\end{equation}
with constraints of the form
\begin{align}
    \phi &= p_t - \bpi \cdot \bv  &  \phi^1&= \Vert\bpi\Vert^2 - p_0^2 &  \bm{\phi}^2&= \p_v \,.
\end{align}
The Lagrangian \eqref{eq:Lcan6} gives the following dynamical
equations of motion:
\begin{subequations}
\begin{align}
\dot{t}&=u & \dot{p}_{t}&=0 & \dot{\boldsymbol{x}}&=2u_{1}\boldsymbol{\pi}-u\boldsymbol{v} \label{eq:EOM6_1}\\
\dot{\boldsymbol{\pi}}&=0 & \dot{\boldsymbol{v}}&=\boldsymbol{u}_{2} & \dot{\boldsymbol{p}}_{v}&=u\boldsymbol{\pi} \,. \label{eq:EOM6_2}
\end{align}
\end{subequations}
The preservation in time of the constraints does not result in
secondary constraints. However, it fixes some of the Lagrange
multipliers, indicating the presence of second-class constraints. From
the preservation of $\phi$ and $\boldsymbol{\phi}^{2}$ one finds that
(the conservation of $\phi^1$ does not yield further equations)
\begin{subequations}
\begin{align}
\boldsymbol{u}_{2}\cdot\boldsymbol{\pi} &=0  \label{eq:upi}\\
  u &= 0 \,. \label{eq:u0}
\end{align}
\end{subequations}
There are two interesting properties that can be derived from the
previous equations. From \eqref{eq:EOM6_2} and~\eqref{eq:upi} one
finds the following restriction on the dynamics
\begin{equation}
    \boldsymbol{\pi}\cdot \dot{\boldsymbol{v}}=0 \,.
\end{equation}
Consequently, the acceleration $\dot{\boldsymbol{v}}$ in the direction
of the momentum $\boldsymbol{\pi}$ must vanish, and the acceleration
in the direction of the plane orthogonal to the momentum will be part
of the gauge freedom (since $\dot{\boldsymbol{v}}=\boldsymbol{u}_{2}$,
where the transverse component of  $\boldsymbol{u}_{2}$ with respect to $\boldsymbol{\pi}$ is arbitrary). This property will play a
key role in the mobility restriction of planons that will be studied
in~\cite{Planons:2024}.

The second important property that can be derived from~\eqref{eq:u0}
is that the equation describing the evolution of the time variable
becomes
\begin{equation}
\dot{t}=0 \,.
\end{equation}
This means that this type of Galilean particle does not evolve in the
physical time $t$, and the orbit is instantaneously defined at a
certain fixed value of $t$. Indeed, it is not possible to choose a
``gauge fixing'' of the form $t=\tau$ as in the previous cases.

Let us now examine the structure of the constraints in more detail.
The second-class constraints are given by
\begin{equation}
  \phi=p_t-\boldsymbol{\pi}\cdot\boldsymbol{v}\,\qquad\qquad\qquad\chi:=\boldsymbol{\pi}\cdot\boldsymbol{p}_{v} \,.
\end{equation}
In particular, its Poisson bracket yields
\begin{equation}
  \left\{ \phi,\chi\right\} = - \Vert \boldsymbol{\pi} \Vert^2 = -p_{0}^{2} \,.
\end{equation}
On the other hand, the first-class constraints are given by 
\begin{equation}
  \left\Vert \boldsymbol{\pi}\right\Vert ^{2}-p_{0}^{2} \approx 0 \,\qquad\qquad\qquad \boldsymbol{p_{v}}^{\perp}:=\boldsymbol{p_{v}}-\frac{1}{p_{0}^{2}}\left(\boldsymbol{\pi}\cdot\boldsymbol{p_{v}}\right)\boldsymbol{\pi}\approx0 \,.
\end{equation}
Note that there are 14 canonical variables, 3 first-class constraints
and 2 second-class constraints. Thus, the number of independent
variables is $14-2\times3-2=6$, which precisely coincides with the
dimension of the orbit.

If we impose the second-class constraints to be strongly equal to
zero, $\phi=\chi=0$, and if in addition we solve the first-class
constraint $\boldsymbol{\pi_{v}}^{\perp}=0$, together with the gauge
fixing condition $\boldsymbol{v}^{\perp}=0$, then the Lagrangian takes
the form
\begin{equation}
L=p_{0}v^{L}\dot{t}+\boldsymbol{\pi}\cdot\dot{\boldsymbol{x}}-\eta_{1}\left(\left\Vert \boldsymbol{\pi}\right\Vert ^{2}-p_{0}^{2}\right),
\end{equation}
where $v_{L}:=\frac{\boldsymbol{\pi}}{p_{0}}\cdot\boldsymbol{v}$ is
the longitudinal component of the velocity.

\subsubsection{Orbit $\#7$} 

The orbit representative for this case is given by
\begin{equation} 
\alpha=\sM(0,0,\boldsymbol{p},\boldsymbol{k},\boldsymbol{0})\,,\label{eq:rep_orbit7}
\end{equation}
where $\Vert\p\Vert = p_0 > 0$, $\Vert\boldsymbol{k}\Vert = k_0 > 0$
and $\Vert\p\times \boldsymbol{k}\Vert=p_0k_0>0$. The Lagrangian is
obtained by using the representative~\eqref{eq:rep_orbit7}
in~\eqref{eq:mc-form-alpha-action}
\begin{equation}
L=\left(R\boldsymbol{p}\cdot\boldsymbol{v}\right)\dot{t}+\left(R\boldsymbol{p}\right)\cdot\dot{\boldsymbol{x}}+\left(R\boldsymbol{k}\right)\cdot\dot{\boldsymbol{v}}\,.
\end{equation}
The Lagrangian in canonical form can then be written as 
\begin{equation}
    L_\text{can}[t,p_t,\x,\bpi,\bv,\p_v,u,u_1,u_2,u_3] = p_t \dot t + \bpi \cdot \dot \x + \p_v \cdot \dot \bv - u \phi - u_i \phi^i \,,
\end{equation}
with
\begin{equation}
    \phi = p_t - \bpi \cdot \bv \,,\quad \phi^1 = \Vert\bpi\Vert^2 - p_0^2 \,,\quad \phi^2 = \Vert\p_v\Vert^2 - k_0^2 \,,\quad \phi^3 = \bpi \cdot \p_v \,.\label{eq:constraintsorbit7}
\end{equation}
The corresponding dynamical equations of motion are given by 
\begin{align}
  \dot{t} & =u & \dot{p}_{t}&=0 &\dot{\boldsymbol{\pi}}&=0 \,,\\
  \dot{\boldsymbol{x}} & =-u\boldsymbol{v}+2u_{1}\boldsymbol{\pi}+u_{3}\boldsymbol{p}_{v} & \dot{\boldsymbol{v}} &= 2u_{2}\boldsymbol{p}_{v}+u_{3}\boldsymbol{\pi} & \dot{\boldsymbol{p}}_{v}&=u\boldsymbol{\pi} \,.
\end{align}
The preservation of the constraints under time evolution does not
generate secondary constraints. Nevertheless, some of the Lagrange
multipliers are determined by the equations of motion, indicating the
presence of second-class constraints. In particular, the preservation
in time of $\phi$ and $\phi_{3}$ implies that
\begin{equation}
    u=u_3=0 \,. \label{eq:Lagmultzero}
\end{equation}
Indeed, it is straightforward to show that the set
$\left(\phi,\phi_3\right)$ defines second-class constraints with a
non-vanishing Poisson bracket given by
\begin{equation}
    \left\{ \phi,\phi_{3}\right\} =-p_{0}^{2} \,.
\end{equation}
On the other hand, the first-class constraints are given by
\begin{equation}
    \left\Vert \boldsymbol{\pi}\right\Vert ^{2}-p_{0}^{2}\approx0\,\qquad\qquad\qquad\left\Vert \boldsymbol{p}_{v}\right\Vert ^{2}-k_{0}^{2}\approx 0 \,.
\end{equation}
There are 14 canonical variables, 2 first-class constraints and 2
second-class constraints. Consequently a direct counting of the
degrees of freedom gives $14-2\times2-2=8$ independent variables. This
is precisely the dimension of the orbit.

Using~\eqref{eq:Lagmultzero} the dynamical equations of motion can be
rewritten as follows
\begin{subequations}
\begin{align}
\dot{t} & =0 & \dot{p}_{t}&=0 & \dot{\boldsymbol{\pi}}&=0 \,,\\
\dot{\boldsymbol{x}} & =2u_{1}\boldsymbol{\pi} & \dot{\boldsymbol{v}}&= 2u_{2}\boldsymbol{p}_{v} & \dot{\boldsymbol{p}}_{v}&=0 \,.
\end{align}
\end{subequations}
In particular, like in orbit $\#6$, the condition $\dot{t}=0$ implies
that this specific type of Galilean particle does not evolve in the
physical time $t$. It is defined at a given instant of time and
relates simultaneous events.

Additionally, one finds the following conditions:
\begin{equation}
\boldsymbol{\pi}\cdot\dot{\boldsymbol{v}}=0\qquad\qquad\qquad\boldsymbol{p}_{v}\cdot\dot{\boldsymbol{x}}=0 \,.
\end{equation}
Alongside the equations of motion, these conditions imply that the
component of the acceleration $\dot{\boldsymbol{v}}$ parallel to the
momentum $\boldsymbol{\pi}$ vanishes, the component of the
acceleration orthogonal to $\boldsymbol{\pi}$ and $\boldsymbol{p}_{v}$
also vanishes, while the component perpendicular to $\boldsymbol{\pi}$
and parallel to $\boldsymbol{p}_{v}$ is pure gauge. Additionally, the
component of $\dot{\boldsymbol{x}}$ parallel to $\boldsymbol{p}_{v}$
vanishes, the one orthogonal to $\boldsymbol{p}_{v}$ and
$\boldsymbol{\pi}$ also vanishes, while the component orthogonal to
$\boldsymbol{p}_{v}$ and parallel to $\boldsymbol{\pi}$ is pure gauge.
This type of restriction in the dynamics plays a crucial role in the
study of planons\cite{Planons:2024}.

To solve the second-class constraint $\phi_{3}=0$ one can decompose
$\boldsymbol{p}_{v}$ into its longitudinal and transverse components
relative to $\boldsymbol{\pi}$
\begin{equation}
  \boldsymbol{p}_{v} = \boldsymbol{p}_{v}^{L}+\boldsymbol{p}_{v}^{\perp} \,,
\end{equation}
where 
\begin{equation}
  \boldsymbol{p}_{v}^{L}=\hat{\boldsymbol{n}}\cdot\boldsymbol{p}_{v} \,,\quad\boldsymbol{p_{v}}^{\perp}:=\boldsymbol{p_{v}}-\left(\hat{\boldsymbol{n}}\cdot\boldsymbol{p_{v}}\right)\hat{\boldsymbol{n}} \,,
\end{equation}
with
\begin{equation}
  \hat{\boldsymbol{n}}=\frac{\boldsymbol{\pi}}{p_{0}} \,.
\end{equation}
Hence, the second-class constraints $\left(\phi,\phi_3\right)$  in Eq.~\eqref{eq:constraintsorbit7} are solved by imposing that
\begin{equation}
  p_{t}=p_{0}v^{L}\,,\quad\boldsymbol{p}_{v}^{L}=0 \,.
\end{equation}
where $\boldsymbol{v}^{L}=\hat{\boldsymbol{n}}\cdot\boldsymbol{v}$.
Therefore, the Lagrangian reduces to
\begin{equation}
  L\left[t,\x,\bpi,\bv,\p_v,u_1,u_2\right]=p_{0}v^{L}\dot{t}+\boldsymbol{\pi}\cdot\dot{\boldsymbol{x}}+\boldsymbol{p}_{v}^{\perp}\cdot\dot{\boldsymbol{v}}-u_{1}\left(\left\Vert \boldsymbol{\pi}\right\Vert ^{2}-p_{0}^{2}\right)-u_{2}\left(\left\Vert \boldsymbol{p}_{v}^{\perp}\right\Vert ^{2}-k_{0}^{2}\right).
\end{equation}

\section{A geometrical approach to Galilei particle dynamics}
\label{sec:particle-dynamics}

In the previous section we have analysed the dynamics described by the
action~\eqref{eq:action-for-alpha} associated to the coadjoint orbits
$\eO_\alpha$.  In this section we will briefly outline a geometrical
approach to studying the dynamics.  The starting point is the action 
\eqref{eq:action-for-alpha}, which we can analyse for general
$\alpha$.  As shown, e.g., in
\cite[Appendix~A.4]{Figueroa-OFarrill:2023vbj}, its extrema are given
by curves $g(\tau) = g_0 c(\tau)$, where $g_0\in G$ and $c : I \to G_\alpha$ is an
arbitrary curve in the stabiliser of $\alpha$.  Under the orbit map 
$\pi_\alpha: G \to \eO_\alpha$, the curve is sent to the constant
$\Ad_{g_0}^*\alpha$, which defines a point in $\eO_\alpha$.

We may interpret $\Ad_{g_0}^*\alpha$ as the momentum of a particle
moving in any homogeneous spacetime of $G$ whose trajectory is given
by composing the curve $g(\tau)$ with the orbit map associated to the
spacetime.  Let $M$ denote a homogeneous spacetime and let $o \in M$
be a choice of origin.  Then $M$ is $G$-equivariantly diffeomorphic to
$G/G_o$, with $G_o$ the stabiliser subgroup of $o$.  We let $\pi_o : G
\to M$ denote the associated orbit map.  Let $K:= G_\alpha \cap G_o$
and consider the homogeneous space $G/K$. The following commutative
diagram summarises the relations between these spaces:
\begin{equation}
  \label{eq:spaces}
  \begin{tikzcd}
    & G \arrow[d,"\pi"] \arrow[ldd,bend right=30,swap, "\pi_\alpha"] \arrow[rdd,bend left=30,"\pi_o"] & \\
    & G/K \arrow[ld] \arrow[rd] & \\
    \eO_\alpha & & M
  \end{tikzcd}.
\end{equation}
In particular, the identity coset in $G/K$ maps to both $\alpha \in
\eO_\alpha$ and $o \in M$.  Let $\pi: G \to G/K$ denote the orbit map
relative to the identity coset.  Let $g(\tau)$ be a curve in $G$ which
extremises the action~\eqref{eq:action-for-alpha}.  We have seen that
it maps to a point in $\eO_\alpha$, which can be interpreted as the
momentum of the particle trajectory $\pi_o(g(\tau))$ in $M$.  We can work
this out by first considering $\pi(g(\tau))$ as a trajectory in $G/K$ and
then mapping that trajectory to $M$.  This amounts to writing
\begin{equation}
  g(\tau) = g_0 \gamma(\tau) k(\tau),
\end{equation}
where $\tau \mapsto k(\tau)$ is a curve in $K$ and $\tau \mapsto \gamma(\tau)$
depends on a choice of coset representative for $G_\alpha/K$.  Then
the particle trajectory on $M$ is simply $g_0 \gamma(\tau)\cdot o$.  The
action of $g_0$ is a global $G$-transformation which amounts to a
change of ``inertial frame'' ($G$ is the relativity group, after all),
so that to understand the particle trajectory on $M$ all we need to do
is to understand $\gamma(\tau)\cdot o$.

Table~\ref{tab:coadjoint-orbits} lists the stabiliser subgroups
$G_\alpha$ for each coadjoint orbit.  Galilei spacetime is described
by a Klein pair $(\g,\g_o)$ with $\g_o = \left<L_{ab},B_a, M\right>$.

It is a simple matter to list generators of $\g_\alpha$ and $\g_\alpha
\cap\g_o$, as well as its complementary space $\fm$ in $\g_\alpha$;
that is, $\g_\alpha =  \fm \oplus (\g_\alpha\cap \g_o)$.  These results are
summarised in Table~\ref{tab:stabilisers}.

\begin{table}[h]
  \centering
    \caption{Stabilisers associated to Galilei particle dynamics}
    \label{tab:stabilisers}
  \begin{adjustbox}{max width=\textwidth}
    \begin{tabular}{l|*{4}{>{$}c<{$}}}
      \# & \alpha =\sM(m,E,\p,\bk,\bj) \in \g^* & \g_\alpha & \g_\alpha \cap \g_o & \fm\\
      \toprule\rowcolor{blue!7}
      1& \sM(m_0,E_0,\bzero,\bzero,\bzero) & \left<M,H,L_a\right> & \left<M,L_a\right> & \left<H\right> \\
      2& \sM(m_0,E_0,\bzero,\bzero,s\be_3) & \left<M,H,L_3\right> & \left<M,L_3\right> & \left<H\right> \\\rowcolor{blue!7}
      3& \sM(0,E_0,\bzero,\bzero,\bzero) & \g & \left<M, B_a, L_a\right> & \left<H,P_a\right>\\
      4& \sM(0,E_0,\bzero,\bzero,s\be_3) & \left<M,H,P_a,B_a,L_3\right> & \left<M, B_a, L_3\right> & \left<H,P_a\right>\\\rowcolor{blue!7}
      5& \sM(0,E_0,\bzero,k_0 \be_3, h\be_3) & \left<M,H,P_a,B_3,L_3\right> & \left<M, B_3, L_3\right> & \left<H,P_a\right>\\
      6& \sM(0,0,p_0\be_3,\bzero,\bzero) & \left<M,P_3,B_1,B_2,L_3\right> & \left<M, B_1,B_2, L_3\right> & \left<P_3\right>\\\rowcolor{blue!7}
      7& \sM(0,0,p_0\be_3,k_0\be_2,\bzero) & \left<M,P_3,B_2\right> & \left<M, B_2\right> & \left<P_3\right>\\
      \bottomrule
    \end{tabular}
  \end{adjustbox}
\end{table}

From this table and in particular from $\fm$, we can deduce the
following, which are in agreement with the analysis in
Section~\ref{sec:action-galil}:
\begin{itemize}
\item For the massive orbits (those of types $\#1,2$), Galilei
  particles can be chosen not to move in space.  This may sound
  surprising, but remember that all statements here are modulo the
  action of the relativity group.  In this case, this simply means
  that any motion in space is an artefact of the choice of inertial
  frame; or in other words, that we can always boost to the rest
  frame.
\item For massless orbits of types $\#3,4,5$, there is no rest frame
  and motion in both space and time is physical.
\item Finally, for massless orbits of type $\#6,7$, Galilei particles
  do not evolve in time: their trajectories instead relate
  simultaneous events.
\end{itemize}

It may be worth comparing this with the case of Carroll particles
treated in \cite{Figueroa-OFarrill:2023vbj}. We let
$\c = \left<L_a, B_a, P_a, H\right>$ denote the Carroll algebra,
$\c_o = \left<L_a,B_a\right>$ the stabiliser Lie algebra of a point in
Carroll spacetime and $\c_\alpha$ the stabiliser Lie algebra of
$\alpha \in \c^*$.  We again let $\fm$ denote a choice of complement
of $\c_\alpha \cap \c_o$ in $\c_\alpha$.  The results are summarised
in Table~\ref{tab:stabilisers-carroll}.  We see that Carroll particles
with nonzero energy do have a rest frame, which explains why they were
referred to as ``massive'' in \cite{Figueroa-OFarrill:2023vbj}.  They
always evolve in time.  Since $\ad_H^* = 0$ in the Carroll algebra,
$H \in \c_\alpha$ for all $\alpha$ and thus also $H \in \fm$ in all
cases, hence any coadjoint orbit $\eO_\alpha$ contains momenta of
Carroll particles which evolve in time, but orbits with zero energy
also contain momenta of Carroll particles which do not.

\begin{table}[h]
  \centering
    \caption{Stabilisers associated to particle dynamics for Carroll particles}
    \label{tab:stabilisers-carroll}
  \begin{adjustbox}{max width=\textwidth}
    \begin{tabular}{l|*{4}{>{$}c<{$}}}
      \# & \alpha =\sM(\bj,\bk,\p,E) \in \c^* & \c_\alpha & \c_\alpha \cap \c_o & \fm \\
      \toprule\rowcolor{blue!7}
      1& \sM(\bzero,\bzero,\bzero,E_0\neq 0) & \left<H,L_a\right> & \left<L_a\right> & \left<H\right> \\
      2& \sM(s\be_3,\bzero,\bzero,E_0 \neq 0) & \left<H,L_3\right> & \left<L_3\right> & \left<H\right> \\\rowcolor{blue!7}
      3& \sM(\bzero,\bzero,\bzero,0) & \c & \c_o & \left<P_a,H\right> \\
      4& \sM(j\be_3,\bzero,\bzero,0) & \left<L_3,B_a,P_a,H\right> & \left<L_3,B_a\right>& \left<P_a,H\right> \\\rowcolor{blue!7}
      5& \sM(h\be_3,k\be_3,\bzero,0) & \left<L_3,B_3,P_a,H\right> & \left<L_3,B_3\right>& \left<P_a,H\right> \\
      6& \sM(h\be_3,\bzero,p\be_3,0) & \left<L_3,B_a,P_3,H\right> & \left<L_3,B_a\right>& \left<P_3,H\right> \\\rowcolor{blue!7}
      7$_\pm$& \sM(h\be_3,\pm k\be_3,p\be_3,0) & \left<L_3,B_3,P_a-\tfrac{p}{k}B_a,H\right> & \left<L_3,B_3\right>& \left<P_a - \tfrac{p}{k}B_a, H\right> \\
      8 & \sM(\bzero,k\cos\theta\be_3 + k\sin\theta\be_2,p\be_3,0) & \left<B_2-\tfrac{k}{p}\cos\theta P_2, B_3+\tfrac{k}{p}\sin\theta P_2,P_3,H\right> & \left<\sin\theta B_2 + \cos\theta B_3\right>& \left<P_2 - \tfrac{p}{k}(\cos\theta B_2 - \sin\theta B_3), P_3, H\right> \\
      \bottomrule
    \end{tabular}
  \end{adjustbox}
\end{table}

\section{From Poincaré to Galilei particles}
\label{sec:from-poinc-galil}

Let us show how to recover a Galilei particle from the $c \to \infty$
limit of a relativistic one. As a first step, we will show how to
recover the Bargmann algebra from a one-dimensional extended Poincaré algebra. Let us start with the
generators of the Poincaré algebra $L_{ab}$, $K_a$, $T_a$ and $T$
verifying
\begin{equation}
\begin{split}
    [L_{ab}, L_{cd}] &= \delta_{bc} L_{ad} - \delta_{ac} L_{bd} - \delta_{bd} L_{ac} + \delta_{bd} L_{ac} \\
    [L_{ab}, K_c] &= \delta_{bc} K_a - \delta_{ac} K_b \\
    [L_{ab}, T_c] &= \delta_{bc} T_a - \delta_{ac} T_b \\
    [K_a, K_b] &= L_{ab} \\
    [K_a, T_b] &= \delta_{ab} T \\
    [K_a, T] &= T_a \,.
\end{split}
\end{equation}
Since the Bargmann algebra has one additional dimension and since
contractions leave the Lie algebra dimension invariant we need to add
an additional element to the Poincaré algebra. We will call this
trivial central extension $M$ and define the new generators from the
relativistic ones
\begin{equation} \label{eq:Poincare-Bargmann-expansion}
    K_a = c B_a \,,\quad T_a = c P_a \,,\quad T = c^2 M + H \,.
\end{equation}
We will assume that all powers of $c$ appear explicitly. In terms of
the new generators, the Lie brackets of the Poincaré algebra become
\begin{equation}
\begin{split}
    [L_{ab}, L_{cd}] &= \delta_{bc} L_{ad} - \delta_{ac} L_{bd} - \delta_{bd} L_{ac} + \delta_{bd} L_{ac} \\
    [L_{ab}, B_c] &= \delta_{bc} B_a - \delta_{ac} B_b \\
    [L_{ab}, P_c] &= \delta_{bc} P_a - \delta_{ac} P_b \\
    [B_a, B_b] &= \frac{1}{c^2} L_{ab} \\
    [B_a, P_b] &= \delta_{ab} M + \frac{1}{c^2} \delta_{ab} H \\
    [B_a, H] &= P_a \,,
\end{split}
\end{equation}
where in the last bracket, we used the fact that $M$ was a central
element. It is easy to see that the $c \to \infty$ limit reproduces
the Bargmann algebra \eqref{eq:barg-alg-lie-basis} where $M$ is now a
nontrivial central extension.

Let us now analyse the implications for the Casimir elements of the
centrally extended Poincaré algebra. First, the quadratic mass-squared
Casimir of Poincaré now reads
\begin{equation}
    T_\mu T^\mu = T_a T^a - T^2 = c^2 P_a P^a - (c^2 M + H)(c^2 M + H) = c^2 P_a P^a - c^4 M^2 - 2 c^2 M H - H^2 \,.
\end{equation}
Upon rescaling by the appropriate power of $c$, the limit $c \to \infty$ yields
\begin{equation} \label{eq:mass-Poincare-Bargmann}
    \lim_{c \to \infty} \frac{1}{c^4} T_\mu T^\mu = - M^2 \,,
\end{equation}
which we already know to be a Galilei Casimir, since $M$ is central. To gain
additional information, one can subtract this contribution of $M^2$
and go to the sub-leading order in $\frac{1}{c}$. Considering the
following limit
\begin{equation}
    \lim_{c \to \infty} \frac{1}{c^2} \left(T_\mu T^\mu + c^4 M^2\right) = P_a P^a - 2 H M \,,
\end{equation}
we recognise the expression of the quadratic Casimir of the
Bargmann algebra, cf.,~Section~\ref{sec:case-n=3}. The interpretation
of this in terms of the non-relativistic limit of the Poincaré
momentum orbit is that the quadratic Casimir of the Bargmann algebra
sits at sub-leading order in the quadratic Casimir of the Poincaré
algebra (extended by $M$), and can be attained once the divergent
mass contribution is properly removed. Finally, the relativistic
Pauli-Lubanski vector gives, in the $c \to \infty$ limit
\begin{equation}
    \lim_{c \to \infty} \frac{1}{c^2} W_a = \epsilon_{abc} (P^b B^c + L^{bc} M) \,,\quad \lim_{c \to \infty} \frac{1}{c^2} W_0 = 0 \,,
\end{equation}
which is a spatial vector affected by spatial rotations only. Its
norm is a conserved quantity
\begin{equation}
    \lim_{c \to \infty} \frac{1}{c^4} W_\mu W^\mu = \epsilon_{abc} (P^b B^c + L^{bc} M) \epsilon^{ade} (P_d B_e + L_{de} M) \,,
\end{equation}
in agreement with the purely bargmannian analysis of
Section~\ref{sec:coadjoint-orbits-3d}.

One should thus be able to obtain all Galilei from Poincaré orbits, at
the expense of adding a central generator to the Poincaré algebra from
the get-go. Let us illustrate how this works at the level of the
momentum orbit. Expanding the Poincaré energy in the same fashion as the generator of time
translations in \eqref{eq:Poincare-Bargmann-expansion}, we obtain $E_\text{P} = m c^2 + E$ and
the mass-shell condition becomes
\begin{equation}
    -m_0^2 c^2 = \frac{1}{c^2} p_\mu p^\mu = - \frac{1}{c^2} E_\text{P} + \Vert\p\Vert^2 = - c^2 m^2 - 2 E m - \frac{E^2}{c^2} + \Vert\p\Vert^2 \,,
\end{equation}
where $m_0^2$ labels orbits in the one-dimensional extension of Poincaré. Since $M$ is a Casimir, the
associated moment $m$ is conserved along the orbit, and the
$c \to \infty$ limit of $T_\mu T^\mu$ identifies $m^2$ with $m_0^2\,$.
Turning now to the sub-leading contribution
\begin{equation}
    \lim_{c \to \infty} \frac{1}{c^2} \left(p_\mu p^\mu + m_0^2 c^4\right) = \Vert\p\Vert^2 - 2 E m_0 := \mu_0 \,,
\end{equation}
we obtain the second constant along the orbit, which we will denote by
$\mu_0$ and which is identified as the value of the quadratic Casimir
of the Bargmann algebra, $P_a P^a - 2HM$. It can assume any real
value. Finally, the spin part gives rise to the eigenvalue of the
last, quartic Casimir
\begin{equation} \label{eq:Poincare-Bargmann-Pauli-Lubanski}
    \lim_{c \to \infty} \frac{1}{c^4} w_\mu w^\mu = \left\Vert m_0 \bj + \p \times \bk\right\Vert^2 := S_0 \,,
\end{equation}
where $w^\mu$ is the Pauli-Lubanski vector. Note that the right-hand
side is always a non-negative number for unitary representations of the Galilei group, while in the usual
parameterisation for the norm of the Pauli-Lubanski vector for UIRs of
Poincaré, this number is non-positive for massless and massive
orbits, and is given by $-m_0^2 c^4 s_0 (s_0+1)$ with
$s_0 \in \mathbb N$. Nevertheless, starting from irreducible (not
necessarily unitary) representations of the Poincaré group with
central extension, we can choose to write $w_\mu w^\mu$ as $c^4 S_0$ where
$S_0$ is any real number, yielding \eqref{eq:Poincare-Bargmann-Pauli-Lubanski}.

In order to have a matching between the coadjoint orbits of the
Poincaré and the Galilei groups, at least at the level of momentum
orbit, one should study the contractions of massless or massive orbits
of the extended Poincaré group. Tachyonic orbits are problematic,
because in \eqref{eq:mass-Poincare-Bargmann} the eigenvalue of the
Casimir of the right-hand side is always a non-positive number for unitary representations, as was
already noticed by Souriau.

\begin{figure}
  \centering
  \scalebox{0.8}{
\begin{tikzpicture}[scale=1]
  \node at (-1,0) {\includegraphics[width=0.5 \textwidth]{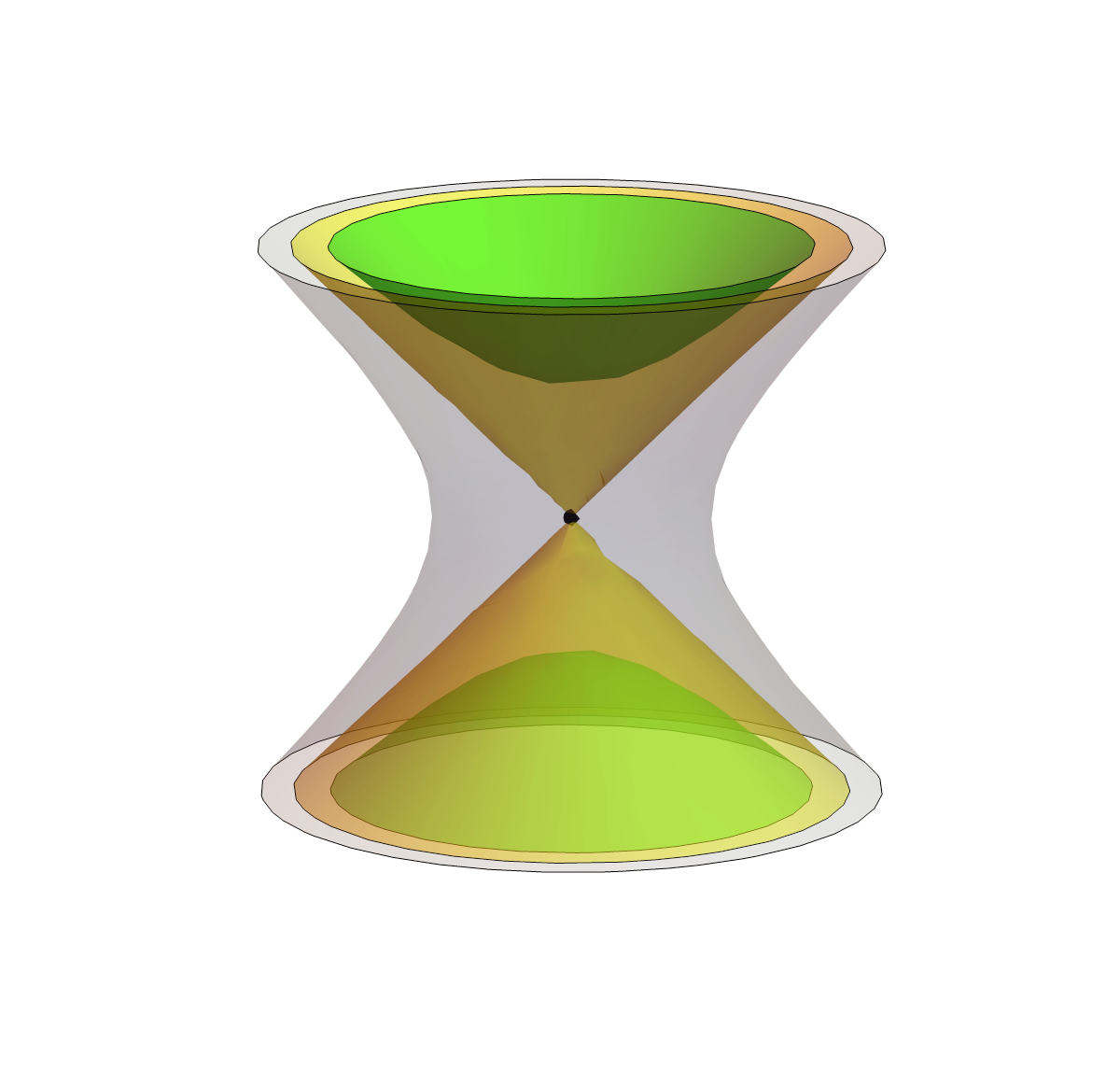}};
  \node at (8,4) {\includegraphics[width=0.5 \textwidth]{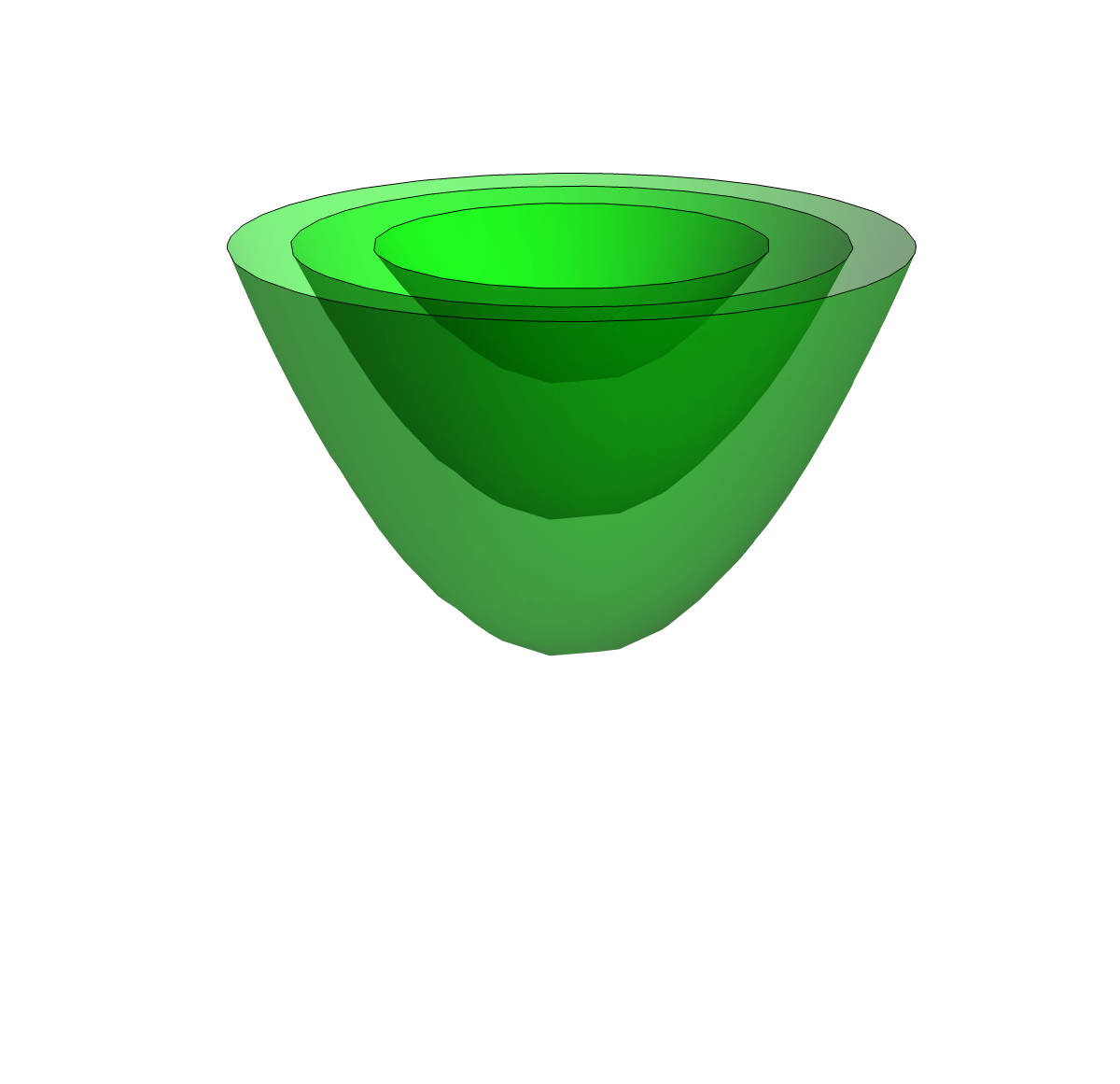}};
  \node at (8,0) {\includegraphics[width=0.5 \textwidth]{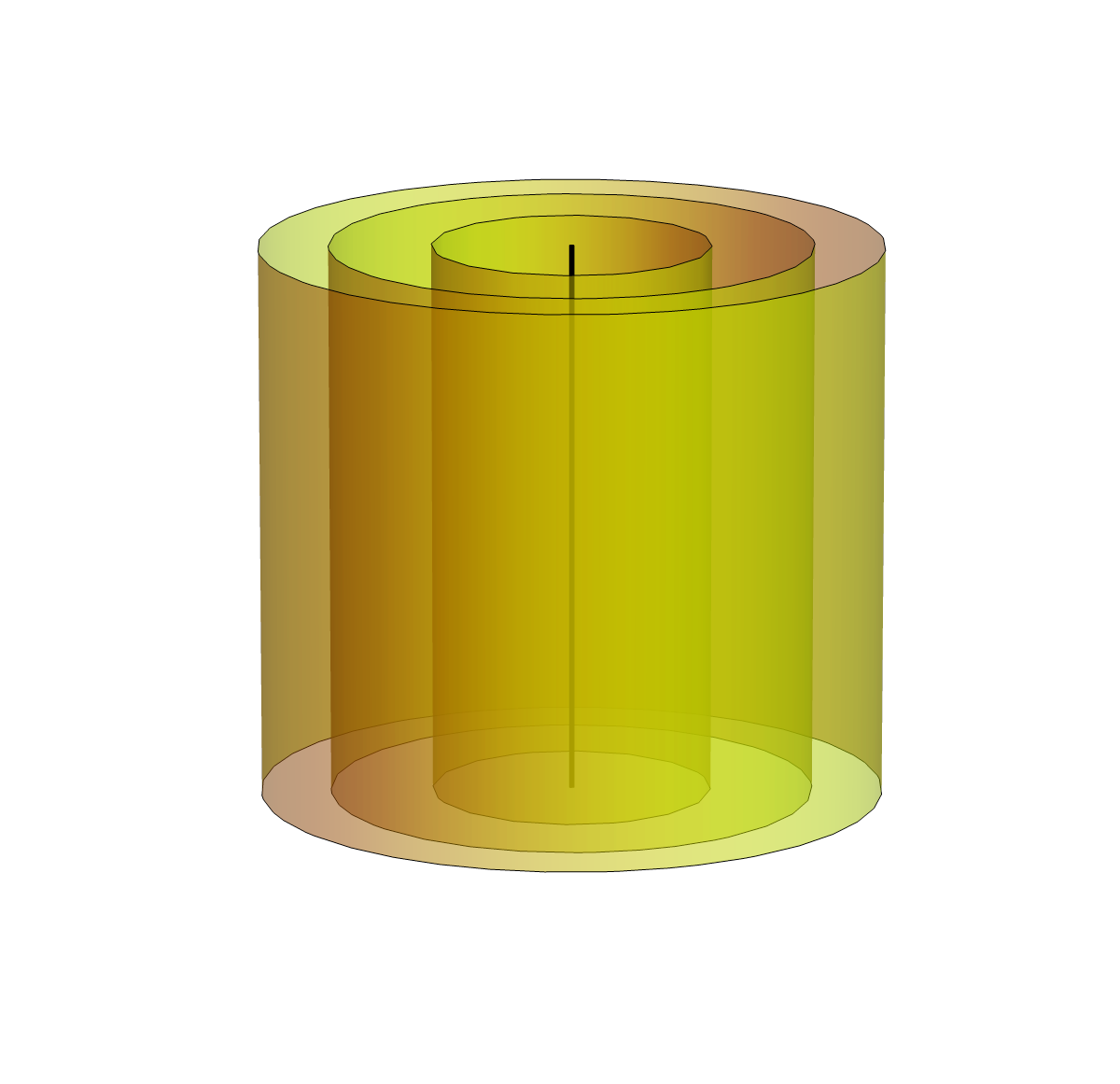}};
  \node at (8,-4) {\includegraphics[width=0.5 \textwidth]{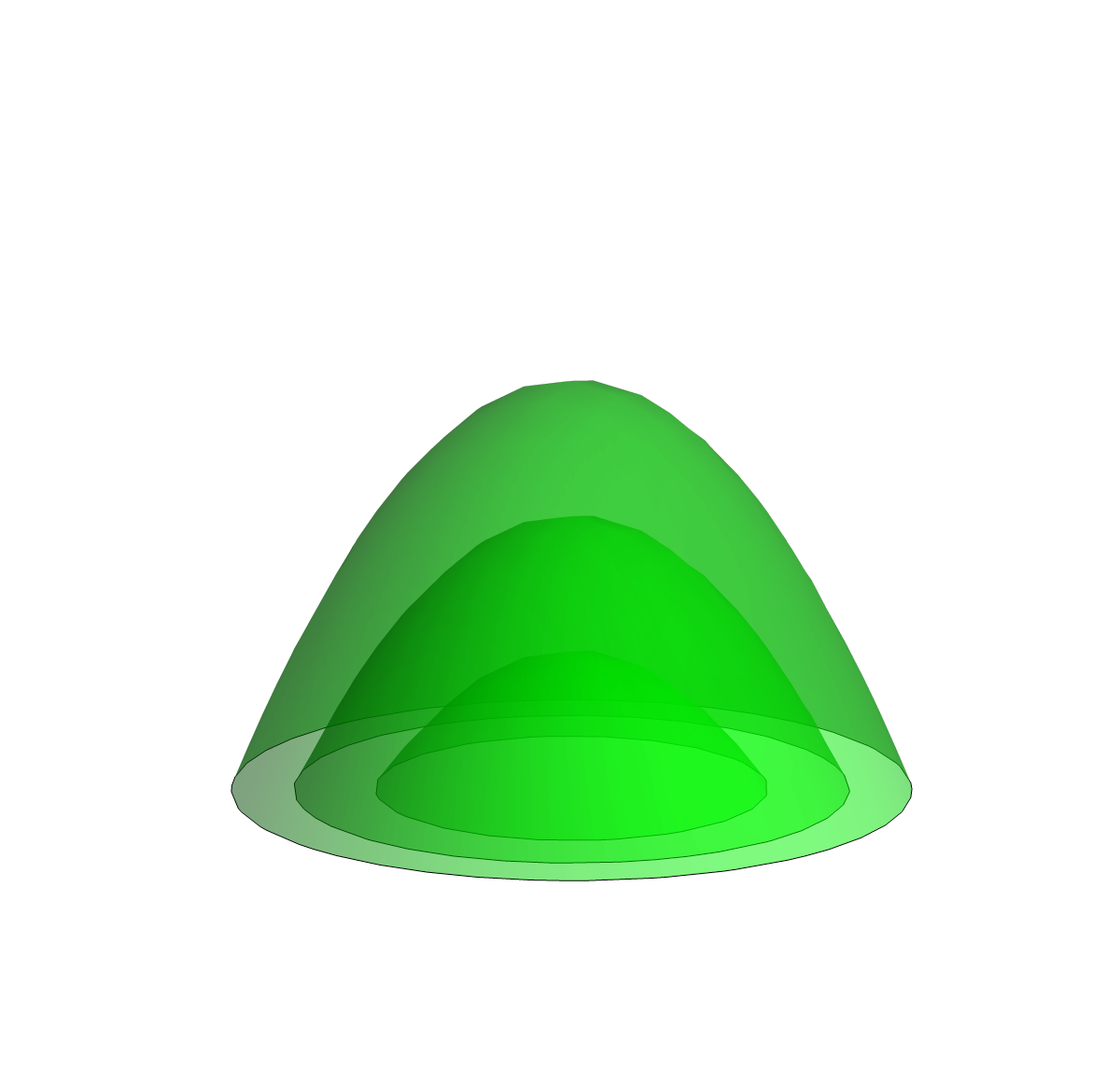}};
  \node at (-1,-7) {Poincaré};
  \node at (8,-7) {Galilei};
   \node[left] at (5.5,4.5) {$m>0$};
   \node[left] at (5.5,0) {$m=0$};
   \node[left] at (5.5,-4.5) {$m<0$};
  \draw[-Stealth,thick] (2.7,-3) -- (2.7,3);
  \node[right] at (2.7,2.9) {$E$};
\end{tikzpicture}
}
\caption{This figure contrasts the energy-momentum orbits $(E,\p)$ of
  the Poincaré (left) and Galilei (right) group. For further details
  we
  refer to Section~\ref{sec:particle-dynamics}, cf., also with Table~\ref{tab:coadjoint-orbits}.\\
  The Poincaré orbits are foliated by hypersurfaces of the form
  $E^{2}- \Vert\p\Vert^2 =m^{2}$. For $m^{2}>0$ this leads to the
  massive orbits with positive and negative energy (green), for $m=0$
  to the massless orbits (yellow) and for $m^{2}<0$ to tachyonic
  orbits (gray).\\
  For the case of Galilei the energy-momentum orbits depend on the
  mass $m$ and we picture three plots for fixed positive, vanishing
  and negative mass. For positive and negative $m$ the energy-momentum
  orbits are hypersurfaces $E-\frac{\Vert\p\Vert^2}{2m}=E_{0}$, where
  $E_{0}$ shifts the parabolas along the energy axis. For vanishing
  mass $m=0$ the Galilei orbits are foliated by cylinders
  $\|\p\|=p_0>0$ and when $\|\p\|=0$ the orbits consist of disjoint
  points $E=E_{0}$ (pictured as a black line).}
    \label{fig:momentum-orb}
\end{figure}

Starting from a massive orbit of extended Poincaré, we obtain massive
orbits $\#1$ or $\#2$ of Galilei, depending on whether $S_0$ is zero or
positive. Starting from a massless Poincaré orbit, we obtain Galilei
orbits $\#3$, $\#4$ and $\#5$ when $\mu_0$ is zero (note that in that
case, $\p$ is itself zero and therefore $S_0$ as well, these orbits
corresponding to different types of Galilei vacua), and orbits $\#6$
or $\#7$ when $\mu_0$ is non-zero (the difference between these last
two orbits is that $S_0$ is zero in the former and non-zero in the
latter). This is depicted in Figure~\ref{fig:momentum-orb}, where massive orbits are depicted in green, and massless ones in yellow. Note that this Figure shows the correspondence between momentum orbits of Poincar\'e (without the one-dimensional extension) to momentum orbits of Bargmann, therefore a single orbit of the former corresponds to a family of orbits of the latter.

\section{Unitary irreducible representations of the Bargmann group}
\label{sec:unit-irred-repr}

In this section we classify the unitary irreducible representations
(UIRs) of (the universal cover of) the Bargmann group via the method
of induced representations.  We are certainly not the first to do
this.  Earlier papers providing (partial) classifications are those of
Inönü--Wigner \cite{MR0050594}, Bargmann \cite{MR0058601},
Lévy-Leblond \cite{MR0154608}, Brennich \cite{MR0466427}, who extended to
representations of the full Bargmann group (including parity and
time-reversal), culminating in the summary of Lévy-Leblond
\cite{MR0275770}.  We will be able to compare these prior
classifications with ours in the end.

\subsection{{$K$}-orbits in {$\ft^*$}}
\label{sec:k-orbits-t-star}

We let $G$ denote the universal cover of the identity component of the
Bargmann group.  It is isomorphic to $K \ltimes T$ where
$K \cong \Spin(3) \ltimes \RR^3$ is the subgroup generated by the
rotations and the boosts and $T \cong \RR^5$ is the abelian normal
subgroup generated by the translations and the central element.  The
action of $K$ on $T$, via conjugation in $G$, differentiates at the
identity to an action of $K$ on $\ft$ and this induces a dual action of
$K$ on $\ft^*$.  We start by choosing $\tau \in \ft^*$ and let
$\eO_\tau$ denote its $K$-orbit.  Letting $K_\tau \subset K$ denote
the stabiliser, we have that $\eO_\tau \cong K/K_\tau$, where the
diffeomorphism is $K$-equivariant.  We may also describe the orbit
somewhat redundantly as $G/(K_\tau \ltimes T)$, where we have
introduced a non-effective action of $G$ on $\eO_\tau$ which shall
nevertheless prove to be very useful.

There are three classes of orbits $\eO_\tau$ of $K$ on $\ft^*$ and
these are summarised in Table~\ref{tab:K-orbits}, where we describe
the orbit and also list the stabiliser of an orbit representative.

\begin{table}[h]
  \centering
    \caption{$K$-orbits in $\ft^*$}
    \label{tab:K-orbits}
  \begin{adjustbox}{max width=\textwidth}
    \begin{tabular}{*{3}{>{$}l<{$}}}
      \tau =(m,E,\p) \in \ft^* & \eO_\tau & K_\tau\\\toprule\rowcolor{blue!7}
     (m,E,\bzero)_{m\neq 0, E\in\RR} & \{(m, E - \tfrac12 m \|\bv\|^2, -m \bv) \mid \bv \in \RR^3\} \cong \RR^3 & \Spin(3) \\
     (0,E,\bzero)_{E\in\RR} & \{(0,E,\bzero)\} & K \\\rowcolor{blue!7}
     (0,0,\p)_{\p \neq \bzero} & \{(0,R\p \cdot \bv, R\p) \mid \bv\in\RR^3, R\in\Spin(3)\} \cong S^2_{\|\p\|} \times \RR & \{(\bv,R) \mid R\p = \p, \bv \perp \p\} \cong \Spin(2)\ltimes \RR^2 \\
      \bottomrule
    \end{tabular}
  \end{adjustbox}
\end{table}

\subsection{Invariant measures}
\label{sec:invar-measures}

Ignoring the point-like orbits, we now show that both types of
three-dimensional orbits admit $K$-invariant measures.  A
$K$-invariant measure is given by integrating a $K$-invariant
nowhere-vanishing $3$-form.  By the holonomy principle, $K$-invariant
nowhere-vanishing $3$-forms on $\eO_\tau$ are in one-to-one
correspondence with nonzero $K_\tau$-invariant elements in
$\wedge^3\fk_\tau^0$, where $\fk_\tau^0 \subset \fk^*$ is the
annihilator of $\fk_\tau$.

Let $L_i, B_i$ denote a basis for $\fk$ and let $\lambda^i, \beta^i$
denote the canonical dual basis for $\fk^*$.  From the Lie brackets of
$\fk$ in this basis
\begin{equation}
  [L_i,L_j] = \epsilon_{ijk} L_k, \qquad [L_i, B_j] = \epsilon_{ijk}
  B_k \qquad\text{and}\qquad [B_i, B_j] = 0,
\end{equation}
we can work out the action of $\fk$ on $k^*$:
\begin{equation}
  \label{eq:coadjoint-k}
  \begin{aligned}
    L_i \cdot \lambda^j &= \epsilon_{ijk} \lambda^k\\
    L_i \cdot \beta^j &= \epsilon_{ijk} \beta^k
  \end{aligned}
  \qquad\qquad
  \begin{aligned}
    B_i \cdot \lambda^j &= 0\\
    B_i \cdot \beta^j &= \epsilon_{ijk} \lambda^k.
  \end{aligned}
\end{equation}

For the orbit with representative $\tau = (m,\frac1{2m}E,\bzero)$,
with $m\neq 0$, the stabiliser Lie algebra $\fk_\tau$ is spanned by the
$L_a$, so that its annihilator $\fk_\tau^0$ is spanned by the
$\beta^a$.  It follows from equation~\eqref{eq:coadjoint-k} that
$\tfrac16 \epsilon_{ijk}\beta^i \wedge \beta^j \wedge \beta^k =
\beta^1 \wedge \beta^2 \wedge \beta^3 \in \wedge^3\fk_\tau^0$ is
$\fk_\tau$-invariant, but since $K_\tau$ is connected, it is also
$K_\tau$-invariant.  We can determine the corresponding $K$-invariant
volume form on the orbit relative to a chart, by choosing a coset
representative $\sigma: \eO_\tau \to K$, with $\sigma(\p) \in K$ such
that $\sigma(\p)\cdot \tau = (m,\frac1{2m}(E - \|\p\|^2, \p)$.  A
possible choice is $\sigma(\p) = \exp(-\tfrac1m \p \cdot \bB)$.  The
pull-back via $\sigma$ of the the left-invariant Maurer--Cartan
one-form on $K$ is given by
\begin{equation}
  \sigma^{-1} d\sigma = \tfrac1m \p \cdot \bB,
\end{equation}
and hence evaluating $\beta^1 \wedge \beta^2 \wedge \beta^3$ on
$\sigma^{-1}d\sigma$ gives the volume form
\begin{equation}
  \dvol = \tfrac1{m^3} dp^1 \wedge dp^2 \wedge dp^3.
\end{equation}
The action of $e^{\bv \cdot \bB} R \in K$ on these coordinates is
calculated by acting on the coset representative $\sigma(\p)$:
\begin{equation}
  e^{\bv \cdot \bB} R \sigma(\p) = e^{\bv \cdot \bB} \sigma(R \p) R =
  \sigma(R\p - m \bv) R,
\end{equation}
so that $\p \mapsto R\p - m \bv$.  This is a euclidean transformation
under which the volume form $\dvol$ is clearly invariant.

For the orbit with representative $\tau = (0,0,\p)$ with $\p =
(0,0,p)$, say, the Lie algebra $\fk_\tau$ of the stabiliser $K_\tau$
is spanned by $L_3,B_1,B_2$, so that its annihilator is spanned by
$\lambda^1, \lambda^2, \beta^3$.  From equation~\eqref{eq:coadjoint-k}
we see that $\lambda^1 \wedge \lambda^2 \wedge \beta^3$ is 
$K_\tau$-invariant.  Under the diffeomorphism $\eO_\tau \cong S^2
\times \RR$, we will see below that the invariant measure is the
product of the measure defined by the area form of the round metric on
$S^2$ and the translationally invariant measure on $\RR$.

\subsection{Inducing representations}
\label{sec:induc-repr}

We induce UIRs of $G$ from UIRs of $K_\tau$ as (square-integrable)
sections of homogeneous vector bundles over $\eO_\tau$.
Square-integrability is defined relative to a $K$-invariant measure on
the orbit, as described in the previous section.

For the three-dimensional orbits with stabiliser $\Spin(3)$, every
UIR is isomorphic to some $V_s$, the complex spin-$s$ representation
of $\Spin(3)$, for $2s \in \{0,1,2,\dots\}$.

For the point-like orbits, the inducing representations are
representations of the euclidean group $K$, which is itself isomorphic
to the semidirect product $\Spin(3) \ltimes \RR^3$ with an abelian
normal subgroup.  We may apply the method of induced representations
to the euclidean group itself.  This was done in
\cite[Section~3.3.1]{Figueroa-OFarrill:2023qty}, for instance, in the
context of the UIRs of the Carroll group.  There are two kinds of UIRs
of $\Spin(3) \ltimes \RR^3$:
\begin{itemize}
\item the complex spin-$s$ representation $V_s$ of $\Spin(3)$, for $2s \in
  \{0,1,2,\dots\}$ with the abelian normal subgroup acting trivially;
\item and the square-integrable sections of the line bundle
  $\cO(n)$ over $\CP^1$ for any $n \in \ZZ$.
\end{itemize}

Finally, for the three-dimensional orbits with $\Spin(2) \ltimes \RR^2$
stabilisers, the possible UIRs can be read off from
\cite[Section~3.3.2]{Figueroa-OFarrill:2023qty}, which considers a
trivial central extension of this group, by ignoring the central
extension.  We find that there are two possible UIRs of $\Spin(2)
\ltimes \RR^2$:
\begin{itemize}
\item one-dimensional representations $\CC_n$ of $\U(1)\cong\Spin(2)$
  with the normal subgroup $\RR^2$ acting trivially;
\item and the square-integrable spinor fields on the
  circle $L^2(S^1,\Sigma_\pm)$ with $\Sigma_+$ (resp. $\Sigma_-$) the
  spinor bundle corresponding to the Ramond (resp. Neveu--Schwarz)
  spin structure on the circle.
\end{itemize}

\subsection{Induced representations}
\label{sec:induced-reps}

The induced representations are carried by square-integrable (with
respect to a $K$-invariant measure) sections of homogeneous vector
bundles over $\eO_\tau$ associated to the inducing representations
just described.  Presumably, the induced representations are obtained
by geometrically quantising the coadjoint orbits of the Bargmann group
and one can hazard a correspondence between the class of orbits and
the induced representations, which we summarise in
Table~\ref{tab:orbits-uirreps} and upon which we elaborate below.

\begin{table}[h]
  \centering
    \caption{Coadjoint orbits and UIRs of the Bargmann group}
    \label{tab:orbits-uirreps}
  \begin{adjustbox}{max width=\textwidth}
    \begin{tabular}{>{$}l<{$}l*{5}{>{$}c<{$}}}
    \multicolumn{1}{c}{Class} & \# & \alpha \in \g^* & \eO_\tau & K_\tau & \text{inducing representation of $K_\tau$} & \text{UIR of $G$}\\\toprule\rowcolor{blue!7}
      \Romanbar{II}(s=0,m,E)& 1& (m,E,\bzero,\bzero,\bzero)_{m\neq 0,E\in\RR} & \RR^3 & \Spin(3) & \CC & L^2(\RR^3,\CC_E)\\
      \Romanbar{II}(s\neq 0,m,E) & 2& (m,E,\bzero,\bzero,\bj)_{m\neq0,E\in\RR,\bj\neq\bzero} & \RR^3 & \Spin(3) & V_{s\neq 0} & L^2(\RR^3,V_s\otimes\CC_E)\\\rowcolor{blue!7}
      \Romanbar{I}(s=0,E) & 3& (0,E,\bzero,\bzero,\bzero)_{E\in\RR} & \{(0,E,\bzero)\} & K & \CC & \CC_E\\
      \Romanbar{I}(s\neq 0,E) & 4& (0,E,\bzero,\bzero,\bj)_{E\in\RR,\bj\neq\bzero} & \{(0,E,\bzero)\} & K & V_{s\neq 0} & V_s \otimes \CC_E\\\rowcolor{blue!7}
      \Romanbar{III}(n,k,E) & 5& (0,E,\bzero,\bk,\bj)_{E\in\RR,\bk\times\bj=\bzero,\bk\neq\bzero} & \{(0,E,\bzero)\} & K & L^2(S^2,\cO(n)) & L^2(S^2,\cO(n)\otimes\CC_E)\\
      \Romanbar{IV}(n,p) & 6& (0,0,\p,\bzero,\bzero)_{\p\neq\bzero} & \RR \times S^2_{\|\p\|} & \Spin(2) \ltimes \RR^2 & \CC_n &L^2(\RR \times S^2,\widetilde{\cO}(-n))\\\rowcolor{blue!7}
      \Romanbar{V}_\pm(p,k^\perp) & 7& (0,0,\p,\bk,\bzero)_{\bk\times\p\neq\bzero} & \RR \times S^2_{\|\p\|} & \Spin(2) \ltimes \RR^2 & \mathscr{H}_\pm:=L^2(S^1,\Sigma_\pm) & L^2_\pm(\RR \times S^3, \CC_{p,k^\perp})\\
      \bottomrule
    \end{tabular}
  \end{adjustbox}
  \vspace{1em}
  \caption*{The table lists a representative $\alpha$ of each class of
    coadjoint orbit $\eO_\alpha$, the base $\eO_\tau$ of the fibration
    which describes $\eO_\alpha$ and the little groups
    $K_\tau \subset K$ from which we induce the UIRs of the Bargmann
    group.  In each row we also list the inducing representation of
    $K_\tau$ as well as the induced representation.  The notation
    $\CC_E$ denotes the copy of $\CC$ on which the one-parameter
    subgroup generated by $H$ acts via the character $\chi(e^{tH}) =
    e^{i E t}$ and also the trivial line bundle associated to that
    one-dimensional representation.  Similarly the notation
    $\CC_{p,k^\perp}$ is the copy of $\CC$ on which the nilpotent subgroup
    generated by $\bB,\bP,M$ acts as in
    equation~\eqref{eq:char-N-class-V} below.  The notation $L^2(X,V)$
    means either $L^2$ functions $X\to V$, when $V$ is a vector space,
    or $L^2$ sections of a vector bundle $V$ over $X$.  The notation
    $\widetilde{\cO}(n)$ denotes the homogeneous line bundle over $\RR
    \times S^2$ obtained by pulling back the line bundle $\cO(n)$ over
    $S^2$ via the cartesian projection $\RR \times S^2 \to S^2$.
    Finally, in the last row, $k^\perp>0$ is defined by
    $\|\p\times\bk\| = p k^\perp$, so that it is the norm of the
    component of $\bk$ perpendicular to $\p$.}
\end{table}

\subsubsection{UIRs of class $\emph{\Romanbar{I}}(s,E)$ associated to orbits of types \#3 and 4}
\label{sec:uirs-orbits-3-4}

These are what we could call the vacuum UIRs.  They are induced from
the finite-dimensional UIRs of $K \cong \Spin(3) \ltimes \RR^3$.  They
are labelled by a non-negative half-integer spin $s$ and a real number
$E$.  The underlying Hilbert space $\eH$ is the complex ($2s+1$)-dimensional
spin-$s$ UIR of $\Spin(3)$ where $H$ acts via the character
$\chi(e^{-a_- H}) = e^{-i a_- E}$.  In other words, $g =
\sg(a_+,a_-,\ba,\bv,R)$ acts on $\psi \in \eH$ as
\begin{equation}
  \label{eq:G-action-class-I}
  g \cdot \psi = e^{i E a_-} \rho(R)\psi,
\end{equation}
with $\rho$ the spin-$s$ representation of $\Spin(3)$.  The inner
product is any $\Spin(3)$-invariant hermitian inner product on $\eH$,
which is unique up to scale.  We label these representations
$\Romanbar{I}(s,E)$, with $E \in \RR$ and $2s \in \{0,1,2,\dots\}$.
They are the galilean analogue of the similarly labelled UIRs of the
Carroll group in \cite{Figueroa-OFarrill:2023qty}.

\subsubsection{UIRs of class $\emph{\Romanbar{II}}(s,m,E)$ associated to orbits of types \#1 and 2}
\label{sec:uirs-orbits-1-2}

These are the massive UIRs. They are induced from the UIRs $V_s$ of
$\Spin(3)$, which are labelled by their non-negative half-integer spin
$s$ and the underlying Hilbert space $\eH = L^2(\RR^3,V_s)$ are the
square-integrable functions $\RR^3 \to V_s$ relative to the standard
euclidean measure on $\RR^3$. We take
$\sigma(\p) = e^{-\tfrac1m \p \cdot \bB}$ as coset representative and
define $\psi(\p) = F(\sigma(\p))$ with $F : G \to V_s$ a Mackey
function equivariant under $K_\tau \ltimes T$. The action of
$g = \sg(a_+,a_-,\ba,\bv,R)$ on $\psi \in \eH$ can be worked out as in
the case of massive Carroll UIRs in \cite{Figueroa-OFarrill:2023qty}
and one finds that
\begin{equation}
  \label{eq:G-action-class-II}
  (g \cdot \psi)(\p) = e^{i(m a_+ + E a_- + \ba\cdot \p)}
  \rho(R) \psi(R^{-1}(\p + m \bv)),
\end{equation}
which is unitary relative to the inner product
\begin{equation}
  \label{eq:IP-class-II}
  \left(\psi_1,\psi_2\right) = \int_{\RR^3} d^3 p \left<\psi_1(\p),\psi_2(\p)\right>_{V_s},
\end{equation}
with $\left<-,-\right>_{V_s}$ any $\Spin(3)$-invariant hermitian inner
product on $V_s$.  These UIRs are labelled by $m\neq 0$, $E \in \RR$
and $s$ with $2s \in \{0,1,2,\dots\}$ and denoted
$\Romanbar{II}(s,m,E)$ and are the galilean analogue of the
similarly-labelled massive Carroll UIRs in
\cite{Figueroa-OFarrill:2023qty}.

\subsubsection{UIRs of class $\emph{\Romanbar{III}}(n,k,E)$ associated to orbits of type \#5}
\label{sec:uirs-orbits-5}

Since the $K$-orbit is point-like, the induced representation shares
the underlying Hilbert space with the inducing representation:
$L^2(S^2,\cO(-n))$ as in the Carroll UIRs of class
$\Romanbar{III}'(n,k)$ in \cite{Figueroa-OFarrill:2023qty}.  We can
read off the results from the Carroll case and we find that $g =
\sg(a_+,a_-,\ba,\bv,R)$ acts on $\psi \in \eH$, which we describe a
complex-valued smooth function on the complex plane $\psi(z)$ with $z$
a stereographic coordinate for the sphere, via
\begin{equation}
  \label{eq:G-action-class-III}
  (g \cdot \psi)(z) = e^{i (E a_- + \bv \cdot \bkappa(z))} \left(
    \frac{\eta + \overline\xi z}{|\eta + \overline\xi z|} \right)^{-n}
  \psi\left( \frac{\overline \eta z - \xi}{\eta + \overline\xi z} \right)
\end{equation}
which is unitary under the inner product
\begin{equation}
  \label{eq:IP-class-III}
  \left(\psi_1,\psi_2\right) = \int_{\CC} \frac{2i dz \wedge
    d\zbar}{(1+|z|^2)^2} \overline{\psi_1(z)}\psi_2(z).
\end{equation}
Here $\bkappa(z) = (\kappa_1(z),\kappa_2(z),\kappa_3(z))$ with
\begin{equation}
  \label{eq:kappa-vector}
    \kappa_1(z) = \frac{2k \Re(z)}{1+|z|^2}, \qquad 
    \kappa_2(z) = \frac{2k \Im(z)}{1+|z|^2} \qquad\text{and}\qquad
    \kappa_3(z) = \frac{(|z|^2-1)k}{1+|z|^2},
\end{equation}
with $k = \|\bk\|$, and where $R \in \Spin(3)$ is given under the
isomorphism $\SU(2) \cong \Spin(3)$ by
\begin{equation}
  \label{eq:R-in-SU2}
  R =  \begin{pmatrix}
    \eta & \xi \\ - \overline \xi & \overline \eta
  \end{pmatrix} \in SU(2).
\end{equation}
These UIRs are labelled by $n \in \ZZ$, $k>0$ and $E \in \RR$ and
denoted $\Romanbar{III}(n,k,E)$.  They are analogous to the Carroll
UIRs of class $\Romanbar{III}'(n,k)$ in \cite{Figueroa-OFarrill:2023qty}.

\subsubsection{UIRs of class $\emph{\Romanbar{IV}}(n,p)$ associated to orbits of type \#6}
\label{sec:uirs-orbits-6}

These UIRs are the analogue of the Carroll UIRs of class
$\Romanbar{III}(n,p)$ in \cite{Figueroa-OFarrill:2023qty}, with one
main difference. Here they are carried by square-integrable sections
of a line bundle over the cylinder $\RR \times S^2$ and not over the
sphere as in the Carroll case. Nevertheless we can re-use many of the
calculations in \cite{Figueroa-OFarrill:2023qty}. Here
$\tau = (0,0,\p)$ where $\p = (0,0,p)$ with $p = \|\p\| >0$. The
stabiliser subgroup is $K_\tau \cong \Spin(2) \ltimes \RR^2$ or, more
invariantly, $\Spin(\p^\perp) \ltimes \p^\perp$ and the orbit is given
by
\begin{equation}
  \label{eq:orbit-class-IV}
  \eO_\tau = \left\{ (0,s,\p) \mid s \in \RR, \p \in S^2_p\right\} \cong \RR \times S^2_p.
\end{equation}
We define a coset representative $\sigma: \eO_\tau \to K$ so
that $\sigma(s,z) \cdot (0,\p) = (s,\boldsymbol{\pi}(z))$, where $z$
is a stereographic coordinate on $S^2_p$; that is,
\begin{equation}
  \boldsymbol{\pi}(z) = \frac{p}{1+|z|^2} \left(2 \Re z, 2 \Im z, |z|^2-1\right),
\end{equation}
which lies in $S^2_p \subset \RR^3$.  A possible choice for
$\sigma(s,z)$ is given by
\begin{equation}
  \sigma(s,z) = e^{s/p^2 \boldsymbol{\pi}(z)\cdot \bB} S(z)
  \qquad\text{where}\qquad S(z) = \frac{1}{\sqrt{1 + |z|^2}}
  \begin{pmatrix}
    z & -1 \\ 1 & \zbar
  \end{pmatrix}.
\end{equation}
Notice that we could also write it as
\begin{equation}
  \label{eq:coset-rep-IV}
  \sigma(s,z) = S(z) e^{s/p^2 \p \cdot \bB},
\end{equation}
using that $\boldsymbol{\pi}(z) = S(z)\cdot \p$.
Let $\chi : T \to \U(1)$ be the character associated to $\tau =
(0,0,\p)$ and let $\CC_n$ denote the UIR of $K_\tau$ where $\Spin(2)$
acts with weight $n$ and the translations in $\p^\perp$ act trivially.
We make $\CC_n$ into a UIR of $K_\tau\ltimes T$ with $T$ acting via
$\chi$.  Let $F: G \to \CC_n$ be a ($K_\tau\ltimes T$)-equivariant
Mackey function.  We define $\psi(s,z) = F(\sigma(s,z))$ and we define
the action of $g \in G$ on $\psi$ via $(g \cdot \psi)(s,z) =
F(g^{-1}\sigma(s,z))$, which we re-express in terms of $\psi$ using
equivariance.

Write $g = \sg(a_+,a_-,\ba,\bv,R)  = e^{a_+ M - a_- H + \ba \cdot \bP}
e^{\bv \cdot \bB} R$, with $R \in \SU(2)$ given as in
equation~\eqref{eq:R-in-SU2}, and let us calculate
\begin{align*}
  g^{-1} \sigma(s,z) &= R^{-1} e^{-\bv \cdot \bB} e^{-a_+ M + a_- H - \ba \cdot \bP} S(z) e^{s/p^2 \p \cdot \bB}\\
                     &= R^{-1} S(z) \underbrace{S(z)^{-1}e^{-\bv \cdot\bB} e^{-a_+ M + a_- H - \ba \cdot \bP} S(z)}_{e^{-S(z)^{-1}\bv \cdot \bB} e^{-a_+ M + a_- H - S(z)^{-1}\ba \cdot \bP}} e^{s/p^2 \p \cdot \bB}\\
                     &= R^{-1} S(z) e^{-(S(z)^{-1}\bv - s/p^2 \p) \cdot \bB} \underbrace{e^{-s/p^2 \p \cdot \bB} e^{-a_+ M + a_- H - S(z)^{-1}\ba \cdot \bP} e^{s/p^2 \p \cdot \bB}}_{e^{-(a_+ - s/p^2 S(z)^{-1}\ba\cdot\p - \tfrac12 a_- s^2/p^2)M + a_- H - (S(z)^{-1}\ba + s/p^2 a_- \p)\cdot \bP}}\\
                     &=R^{-1} S(z) e^{-(S(z)^{-1}\bv - s/p^2 \p) \cdot \bB}e^{-(a_+ - s/p^2 S(z)^{-1}\ba\cdot\p - \tfrac12 a_- s^2/p^2)M + a_- H - (S(z)^{-1}\ba + s/p^2 a_- \p)\cdot \bP}.
\end{align*}
Therefore, using that $S(z)^{-1}\ba \cdot \p = \ba \cdot S(z)\p = \ba \cdot \boldsymbol{\pi}(z)$, we have that
\begin{align*}
  F(g^{-1}\sigma(s,z)) &= F(R^{-1} S(z)  e^{-(S(z)^{-1}\bv - s/p^2 \p) \cdot \bB} e^{-(a_+ - s/p^2 S(z)^{-1}\ba\cdot\p - \tfrac12 a_- s^2/p^2)M + a_- H - (S(z)^{-1}\ba + s/p^2 a_- \p)\cdot \bP})\\
                       &= e^{i (\ba \cdot \boldsymbol{\pi}(z) + s a_-)} F(R^{-1} S(z)  e^{-(S(z)^{-1}\bv - s/p^2 \p) \cdot \bB}),
\end{align*}
where we used the equivariance of $F$.  Using equation~(3.42) in
\cite{Figueroa-OFarrill:2023qty} we have that for $R \in \SU(2)$ given
by equation~\eqref{eq:R-in-SU2},
\begin{equation}
  R^{-1}S(z) = \underbrace{S\left(\frac{\overline\eta - \xi}{\eta + \overline\xi
      z}\right)}_{S(R^{-1}z)} \underbrace{\frac{\eta + \overline\xi z}{|\eta + \overline\xi z|}}_{\lambda(S^{-1},z)\in\U(1)},
\end{equation}
where we identify $\U(1)$ with the diagonal matrices in $\SU(2)$, since those matrices stabilise $\p$.  Therefore,
\begin{align*}
    F(R^{-1} S(z) e^{-(S(z)^{-1}\bv - s/p^2 \p) \cdot \bB}) &= F(S(R^{-1}z) \lambda(S^{-1},z) e^{-(S(z)^{-1}\bv - s/p^2 \p) \cdot \bB})\\
    &=  F(S(R^{-1}z) e^{-((S(z)^{-1}\bv)_\parallel - s/p^2 \p) \cdot \bB}\underbrace{\lambda(S^{-1},z) e^{-((S(z)^{-1}\bv)_\perp \cdot \bB}}_{\in K_\tau})
\end{align*}
where we have used that $\U(1)$ preserves $\p$ and have broken up
$S(z)^{-1}\bv$ into a component along $\p$ (and hence preserved by
$\U(1)$) and a component perpendicular to $\p$.  Using equivariance
again, and the fact that $(S(z)^{-1}\bv)_\parallel = S(z)^{-1}\bv
\cdot \p/p^2 \p = \bv \cdot \boldsymbol{\pi}(z) \p/p^2$, we find
\begin{equation}
  F(g^{-1}\sigma(s,z))  = e^{i (\ba \cdot \boldsymbol{\pi}(z) + s a_-)} \lambda(S^{-1},z)^{-n} F(\sigma(s-\bv\cdot\boldsymbol{\pi}(z),R^{-1}z)).
\end{equation}
In summary, the action of $g$ on $\psi(s,z)$ is given by
\begin{equation}
  \label{eq:G-action-class-IV}
  (g \cdot \psi)(s,z) = e^{-i(\ba\cdot\boldsymbol{\pi}(z) + s a_-)}
  \left( \frac{\eta + \overline\xi z}{|\eta + \overline\xi z|}
  \right)^{-n} \psi\left( s - \bv\cdot\boldsymbol{\pi}(z),
    \frac{\overline\eta z - \xi}{\eta + \overline\xi z} \right).
\end{equation}
This representation is unitary relative to the inner product
\begin{equation}
  \label{eq:IP-class-IV}
  \left( \psi_1,\psi_2 \right) = \int_{\RR \times \CC} \frac{2i ds
    \wedge dz \wedge d\zbar}{(1+|z|^2)^2} \overline{\psi_1(s,z)}\psi_2(s,z).
\end{equation}
More invariantly, and as shown in
\cite[Section~3.3.1]{Figueroa-OFarrill:2023qty} for the case of
Carroll UIRs, one can describe the Hilbert space $\eH$ as the
square-integrable sections of the line bundle ${\widetilde\cO}(-n)$
over the cylinder $\RR \times S^2_p$ obtain by pulling back the line
bundle $\cO(-n)$ over $S^2_p$ via the cartesian projection $\RR \times
S^2_p \to S^2_p$.  We denote these UIRs by $\Romanbar{IV}(n,p)$ where
$n \in \ZZ$ and $p>0$.

\subsubsection{UIRs of class $\emph{\Romanbar{V}}_\pm(p,k^\perp)$ associated to orbits of type \#7}
\label{sec:uirs-orbits-7}

These UIRs are the analogue of the Carroll UIRs of class
$\Romanbar{V}_\pm(p,k,\theta)$ in \cite{Figueroa-OFarrill:2023qty}.
Their description is as sections of an infinite-dimensional Hilbert
bundle over the $K$-orbit, but following similar steps as those in
\cite[Section~3.4.3]{Figueroa-OFarrill:2023qty}, they can be seen to
admit a simpler description.

Let $N$ be the nilpotent subgroup of the Bargmann group generated by
$\bB,\bP,M$.  If $m=0$, $N$ acts like an abelian group and its UIRs
are therefore one-dimensional.  We will consider the one-dimensional
UIR $\CC_{p,k^\perp}$ with character $\chi: N \to \U(1)$ given by
\begin{equation}
  \label{eq:char-N-class-V}
  \chi(e^{a_+ M + \ba\cdot\bP + \bv \cdot \bB}) = e^i(\ba\cdot\p + \bv
\cdot \bk), \qquad\text{with  $\p = (0,0,p)$ and $\bk = (0,k^\perp,0)$ with both $p,k^\perp>0$.}
\end{equation}
The homogeneous space $G/N$ is diffeomorphic to $\RR \times S^3$ and
$\chi$ defines a trivial homogeneous line bundle $L_\chi$ over $G/N$,
whose sections can be identified with functions
$\RR \times S^3 \to \CC_{p,k^\perp}$.

Let us choose a coset representative $\sigma:\RR \times S^3 \to G$ for
$G/N$ defined by $\sigma(s,S) = e^{s H} S$, where $S \in \SU(2)$ and
where we have identified $S^3$ with $\SU(2)$.  The UIR of the Bargmann
group is carried by $\eH = L^2(\RR \times S^3,\CC_\chi)$ relative to
the inner product
\begin{equation}
  \label{eq:IP-class-V}
  \left(\psi_1,\psi_2\right) = \int_{\RR \times S^3} ds d\mu(S) \overline{\psi_1(s,S)}\psi_2(s,S),
\end{equation}
where $d\mu$ is a bi-invariant Haar measure on $\SU(2)$.

Let $g = \sg(a_+,a_-,\ba,\bv,R)$ and let us calculate its action on
$\psi \in \eH$.  As usual $\psi(s,S) = F(\sigma(s,S))$, with $F: G \to
\CC_\chi$ a ($K_\tau \ltimes T$)-equivariant Mackey function.  Then
\begin{equation}
  (g \cdot \psi)(s,S) = F(g^{-1} \sigma(s,S)),
\end{equation}
which we must rewrite in terms of $\psi$ using equivariance.  We
calculate
\begin{align*}
  g^{-1} \sigma(s,S) &= R^{-1} e^{-\bv\cdot\bB} e^{-a_+M + a_- H - \ba\cdot \bP} e^{s H} S\\
                     &= R^{-1} e^{-\bv\cdot\bB} e^{(a_- + s) H} e^{-a_+M - \ba\cdot \bP} S\\
                     &= R^{-1} e^{(a_- + s) H}\underbrace{e^{-(a_- + s) H} e^{-\bv\cdot\bB} e^{(a_- + s) H}}_{e^{-\bv\cdot\bB - (a_- + s) \bv\cdot\bP}} e^{-a_+M - \ba\cdot \bP} S\\
                     &= e^{(a_- + s) H} R^{-1} S e^{-S^{-1}\bv\cdot\bB - (a_- + s) S^{-1}\bv\cdot\bP} e^{-a_+M - S^{-1}\ba\cdot \bP}.
\end{align*}
Therefore, using equivariance, we see that
\begin{equation}
  (g\cdot \psi)(s,S) = e^{i(\bv\cdot S\bk + (\ba + (a_- + s) \bv)\cdot S\p)} \psi(s + a_-, R^{-1}S),
\end{equation}
where we have used that, say, $S^{-1}\bv \cdot \bk = \bv \cdot S\bk$.
As in the case of the similar UIRs of the Carroll group, these
representations are not irreducible, because of the action of the
centre of $\SU(2)$.  We define idempotents $\Pi_\pm : \eH \to \eH$ by
\begin{equation}
  \label{eq:projectors-class-V}
  (\Pi_\pm \psi)(s,S) = \tfrac12 \left( \psi(s,S) \pm \psi(s,-S)
  \right).
\end{equation}
Then $\eH = \eH_+ \oplus \eH_-$, with $\eH_\pm$ the image of $\Pi_\pm$,
is an orthogonal decomposition into UIRs of the Bargmann group.
These UIRs are characterised by $p,k^\perp > 0$ and the action of the centre
of $\SU(2)$, which is a sign.  We will denote them by
$\Romanbar{V}_\pm(p,k^\perp)$ by analogy with the Carroll UIRs in
\cite{Figueroa-OFarrill:2023qty}.

\subsection{Comparison with Carroll UIRs}
\label{sec:comp-with-carr}

As we have been mentioning during the description of the Bargmann UIRs
in the previous section, there are certain similarities between the
Bargmann and Carroll UIRs that are worth highlighting.
Table~\ref{tab:comparison-Carroll} summarises these similarities.

\begin{table}[h]
  \centering
    \caption{Comparison with Carroll UIRs}
    \label{tab:comparison-Carroll}
    \setlength{\tabcolsep}{20pt}
    \begin{tabular}{*{2}{>{$}l<{$}}}
    \multicolumn{1}{c}{Carroll UIR in \cite{Figueroa-OFarrill:2023qty}} & \multicolumn{1}{c}{UIR in Table~\ref{tab:orbits-uirreps}}\\\toprule\rowcolor{blue!7}
      \Romanbar{I}(s) & \Romanbar{I}_+(s,E)\\
      \Romanbar{II}(s,m) & \Romanbar{II}(s,m,E)\\\rowcolor{blue!7}
      \Romanbar{III}'(n,k) & \Romanbar{III}(n,k,E)\\
      \Romanbar{III}(n,p) & \Romanbar{IV}(n,p)\\\rowcolor{blue!7}
      \Romanbar{IV}_\pm(n,p,k) & \\
      \Romanbar{V}_\pm(p,k,\theta) & \Romanbar{V}_\pm(p,k\sin\theta)\\\bottomrule
    \end{tabular}
  \vspace{1em}
  \caption*{This table provides a sort of dictionary between the UIRs
    of the Carroll group as determined in
    \cite{Figueroa-OFarrill:2023qty}  and the UIRs of the Bargmann
    group we have just described.  The correspondence is not perfect:
    there are labels in the Bargmann case which simply do not exist in
    the Carroll case and there are Carroll UIRs which have not
    counterpart among the Bargmann UIRs; namely, the (anti)parallel
    helicity representations of Carroll.}
\end{table}

\subsection{Comparison with prior classifications}
\label{sec:comp-with-prior}

The earliest classification of Galilei UIRs is that of Inönü--Wigner
\cite{MR0050594} who restricted themselves to honest (not ray) UIRs of
the Galilei group, despite being aware (citing a private communication
with none other than Bargmann himself!) of the need to consider
projective representations.  Moreover they classify UIRs of the
connected component of the Galilei group, but not of its
simply-connected double cover.  Therefore their list should be
compared with those UIRs with $m=0$ and with integer spin and
helicity.

Restricting to the Galilei group has the technical advantage that the
maximal abelian subgroup is now of larger dimension than in the
Bargmann case.  Letting $G'$ denote the Galilei group, we can write
$G'= K'\ltimes T'$, where $K'\cong \RR \times \SO(3)$ is the connected subgroup
generated by $J_i, H$, whereas $T' \cong \RR^6$ is the abelian
subgroup generated by $B_i,P_i$.  The UIRs of $T'$ are one-dimensional
and defined by characters $\chi : T' \to \U(1)$ with
\begin{equation}
  \label{eq:char-IW}
  \chi( e^{\bv\cdot\bB + \ba \cdot\bP}) = e^{i(\bv\cdot\bk + \ba \cdot \p)},
\end{equation}
for some $(\bk,\p) \in \RR^6$.  The $K'$-action on the characters is
such that
\begin{equation}
  R e^{s H} \cdot (\bk,\p) = (R \bk + s R \p, R\p).
\end{equation}
There are four types of orbits depending on $\tau'=(\bk,\p)$, in
increasing dimension of the stabiliser $K'_{\tau'} \subset K'$ with
the labels as in \cite{MR0050594}:
\begin{enumerate}[label=(\Roman*)]
\item $\tau'= (\bk,\p)$, with $\bk \times \p \neq \bzero$.  Let $p =
  \|\p\|>0$ and $h = \|\p \times \bk\| > 0$, which are the two
  invariants of the orbit.  The stabiliser is trivial and hence the
  orbit is $\RR \times \SO(3)$.
\item $\tau'= (\bk,\p)$, with $\p\neq\bzero$ and $\bk \times \p =
  \bzero$.  Letting $p = \|\p\|>0$, the orbit is now $\RR \times
  S^2_p$ and the stabiliser is $K'_{\tau'} = \left\{R \mid R\p = \p
  \right\} \cong \SO(2)$.
\item $\tau'= (\bk,\bzero)$, with $\bk \neq \bzero$.  The orbit is the
  sphere of radius $k=\|\bk\|>0$. The stabiliser is $K'_{\tau'} =
  \left\{e^{a H} R \mid a\in\RR,\, R\bk = \bk \right\} \cong \RR
\times \SO(2)$.
\item $\tau'= (\bzero,\bzero)$.  This is a point-like orbit with
  stabiliser all of $K'$.
\end{enumerate}
The inducing UIRs of the stabilisers are easy to determine in all
cases:
\begin{enumerate}[label=(\Roman*)]
\item The stabiliser is trivial, so there is the only UIR is the
  trivial one-dimensional representation $\CC$.
\item The stabiliser is $\SO(2)$ whose UIRs are one-dimensional and
  denoted $\CC_n$ with $n \in \ZZ$, which we may identify with the
  helicity.
\item The stabiliser is $\RR \times \SO(2)$, whose UIRs are
  one-dimensional $\CC_n \otimes \CC_e$, with $n \in \ZZ$ and $e \in
  \RR$.
\item The stabiliser is $\RR \times \SO(3)$, whose UIRs are $V_\ell
  \otimes \CC_e$, with the spin $\ell \in \ZZ$ and $e \in \RR$.
\end{enumerate}
It is then easy to compare their classification with ours and we
give the dictionary in Table~\ref{tab:comparison-IW}.

\begin{table}[h]
  \centering
    \caption{Comparison with Inönü--Wigner \cite{MR0050594}}
    \label{tab:comparison-IW}
    \setlength{\tabcolsep}{20pt}
    \begin{tabular}{*{2}{>{$}l<{$}}}
    \multicolumn{1}{c}{UIR in \cite{MR0050594}} & \multicolumn{1}{c}{UIR in Table~\ref{tab:orbits-uirreps}}\\\toprule\rowcolor{blue!7}
      \Romanbar{I}(p,s) & \Romanbar{V}_+(p,s/p)\\
      \Romanbar{II}(m,p) & \Romanbar{IV}(2m,p)\\\rowcolor{blue!7}
      \Romanbar{III}(m,k,e) & \Romanbar{III}(2m,k,e)\\
      \Romanbar{IV}(\ell,e) & \Romanbar{I}(2\ell, e)\\\bottomrule
    \end{tabular}
  \vspace{1em}
  \caption*{This table provides a dictionary between the
    representations of the Galilei group classified in
    \cite{MR0050594} and the ones in our
    Table~\ref{tab:orbits-uirreps}.  The notation in the first column
    is the one adopted in \cite{MR0050594}.  In particular, their $m$
    is not the mass, but an integer helicity.  Besides the massive
    representations, also missing are any Bargmann UIRs where the centre of
    $\SU(2)$ acts nontrivially.  We reiterate that the missing UIRs
    were consciously and explicitly excluded in \cite{MR0050594}.}
\end{table}

The UIRs classified in Lévy-Leblond \cite{MR0154608} and Brennich
\cite{MR0275770}, although expressed in the language of ray
representations of the Galilei group instead of representations of the
Bargmann group, are induced from characters of the abelian subgroup
$T$ generated by $M,H,P_i$ together with a UIR of the stabiliser of
the character, as we have done.  This allows for an easier comparison
than in the case of Inönü--Wigner.  We give the dictionary in
Tables~\ref{tab:comparison-B} and \ref{tab:comparison-LL}.

\begin{table}[h]
  \centering
    \caption{Comparison with Brennich \cite{MR0275770}}
    \label{tab:comparison-B}
    \setlength{\tabcolsep}{20pt}
    \begin{tabular}{*{2}{>{$}l<{$}}}
    \multicolumn{1}{c}{UIR in \cite{MR0275770}} & \multicolumn{1}{c}{UIR in Table~\ref{tab:orbits-uirreps}}\\\toprule\rowcolor{blue!7}
      \Romanbar{I}(s,m,e,\p)& \Romanbar{II}(s,m,e)\\
      \Romanbar{II}(s,p) & \Romanbar{IV}(2s,p)\\\rowcolor{blue!7}
      \Romanbar{III}(s,p,k,e) & \Romanbar{V}_{\pm(s)}(p,k)\\
      \Romanbar{IV}(s,e) & \Romanbar{I}(s,e)\\\rowcolor{blue!7}
      \Romanbar{V}(s,k,e) & \Romanbar{III}(2s,k,e)\\\bottomrule
    \end{tabular}
  \vspace{1em}
  \caption*{This table provides a dictionary between the unitary
    irreducible ray representations of the Galilei group classified in
    \cite{MR0275770} and the UIRs of the Bargmann group in
    Table~\ref{tab:orbits-uirreps}.  The labels $\p$ and $e$ in the
    UIRs of classes $\Romanbar{I}$ and $\Romanbar{III}$, respectively,
    in \cite{MR0275770} are spurious, since they are not actually
    invariants.  They correspond to a choice of inducing
    character.  Also, the label $s$ in Brennich's
    $\Romanbar{III}(s,k,e)$ takes two possible values: $s=0$,
    corresponding to our $\Romanbar{V}_+$, and $s=\tfrac12$,
    corresponding to our $\Romanbar{V}_-$.  Other than those comments,
    there is a bijective correspondence between the UIRs in
    \cite{MR0275770} and the ones in Table~\ref{tab:orbits-uirreps}.
    Of course, Brennich also discusses the UIRs of the full Bargmann
    group, including inversion and time-reversal.}
\end{table}

\begin{table}[h]
  \centering
    \caption{Comparison with Lévy-Leblond \cite{MR0466427}}
    \label{tab:comparison-LL}
    \setlength{\tabcolsep}{20pt}
    \begin{tabular}{*{2}{>{$}l<{$}}}
    \multicolumn{1}{c}{UIR in \cite{MR0466427}} & \multicolumn{1}{c}{UIR in Table~\ref{tab:orbits-uirreps}}\\\toprule\rowcolor{blue!7}
      \Romanbar{I}(p,v)& \Romanbar{V}_+(p,v)\\
      \Romanbar{II}(p,\sigma) & \Romanbar{IV}(\sigma,p)\\\rowcolor{blue!7}
      \Romanbar{III}(E,k,\xi) & \Romanbar{III}(2\xi,k,E)\\
      \Romanbar{IV}(E,\ell) & \Romanbar{I}(\ell,E)\\\rowcolor{blue!7}
      m(U,s) & \Romanbar{II}(s,m,U)\\\bottomrule
    \end{tabular}
  \vspace{1em}
  \caption*{This table provides a dictionary between the unitary
    irreducible ray representations of the Galilei group classified in
    \cite{MR0466427} and the UIRs of the Bargmann group in
    Table~\ref{tab:orbits-uirreps}.  The UIR
    $\Romanbar{V}_-(p,k^\perp)$ in Table~\ref{tab:orbits-uirreps}
    is missing from the list in \cite{MR0466427}, but
    otherwise we are in agreement.  This can be explained by the fact
    that the euclidean group from which one induces the representation
    is actually the double cover of the euclidean group in
    \cite{MR0466427} and hence the ``little group'' mentioned after
    equation (4.14) there is not actually the identity but the order-2
    Galois group of the double cover.}
\end{table}

\section{Galilean field-theoretical realisations}
\label{sec:galil-field-theor}

In this section we will realise some of the UIRs of the Bargmann group
in terms of fields in Galilei spacetime.  This follows the method
explained in \cite[Appendix~A]{Figueroa-OFarrill:2023qty}.

Of the UIRs of the Bargmann group, there are some which admit a
description in terms of (finite-component) fields on Galilei
spacetime.  Galilei spacetime is a homogeneous space of the Bargmann
group diffeomorphic to the space of cosets $G/G_0$, where
$G_0 = K \times Z$ with $Z$ the central subgroup generated by $M$.  As
explained, for example in
\cite[Appendix~A]{Figueroa-OFarrill:2023qty}, the first step in
obtaining such a description is to embed the inducing representation
of $K_\tau$ into a (finite-dimensional) representation of $G_0$.  This
is possible for all inducing representations except those associated
with the coadjoint orbits of classes \#5,7.  Those representations
associated to coadjoint orbits of classes \#3,4 are
finite-dimensional, so presumably they do not admit a nontrivial
description as fields on Galilei spacetime.  Thus we remain with the
UIRs of classes $\Romanbar{II}(s,m,E)$ associated with coadjoint orbits
of types \#1 and 2 and $\Romanbar{IV}(n,p)$ associated with coadjoint
orbits of type \# 6.

\subsection{Massive galilean fields}
\label{sec:mass-galil-fields}

In Section~\ref{sec:uirs-orbits-1-2} we described the momentum-space
description of the UIRs of type $\Romanbar{II}(s,m,E)$ and in this
section we will realise these representations as fields in Galilei
spacetime.

Galilei spacetime is diffeomorphic to the coset space $G/G_0$ with
$G_0= K \times Z$.  Fields on Galilei spacetime are sections of
homogeneous vector bundles associated to representations of $G_0$, so
the first order of business is to \emph{choose} a finite-dimensional
representation of $G_0$ which embeds the inducing representation $V$
of $K_\tau$.  Extending the representation $V_s$ from $K_\tau$ to $K$
is simply a matter of letting the boosts act trivially.  We may also
extend it to a representation of $G_0$ via
\begin{equation}
  \sg(a_+,0,\bzero,\bv,R) \cdot \psi = e^{ima_+} R\cdot \psi.
\end{equation}
Let $V$ denote this representation of $G_0$, sharing the same vector
space with the representation $V_s$ of $K_\tau$.  The action of
$G_0$ on $\eO_\tau$ is such that the central subgroup $Z$ acts
trivially and $K$ acts via euclidean transformations, as seen above.

Next we ``Fourier transform''.  We define $\widehat F : G \to V$
by\footnote{We tacitly restrict to Mackey functions $F$ for which this
  integral converges.}
\begin{equation}
  \widehat F(g) := \int_{\RR^3} d^3p\ \sigma(\p) \cdot F(g\sigma(\p)) =
  \int_{\RR^3} d^3p\ F(g\sigma(\p)),
\end{equation}
since the boost $\sigma(\p) = e^{-\frac1m \p\cdot \bB}$ acts trivially
on $V$.  As shown, for example, in
\cite[Appendix~A]{Figueroa-OFarrill:2023qty}, $\widehat F$ is
$G_0$-equivariant and hence it defines a section of the homogeneous
vector bundle over Galilei spacetime associated with the
representation $V$.

Let $\zeta : G/G_0 \to G$ be a coset representative for Galilei
spacetime, where $\zeta(t,\x) = \exp(tH + \x\cdot \bP)$ and define
$\phi: G/G_0 \to V$ by
\begin{equation}
  \phi(t,\x) = \widehat F(\zeta(t,\x)) = \int_{\RR^3} d^3p\ F(\zeta(t,\x)\sigma(\p)).
\end{equation}
We now calculate
\begin{equation}
  \zeta(t,\x) \sigma(\p) = \sigma(\p) \underbrace{\sigma(\p)^{-1} \zeta(t,\x) \sigma(\p)}_{\in T},
\end{equation}
where, after a quick calculation, we find that
\begin{equation}
  \sigma(\p)^{-1} \zeta(t,\x) \sigma(\p) =\zeta(t, \x + \tfrac1m \p t) e^{\frac1m(\x \cdot \p + \frac1{2m}\|\p\|^2)M}.
\end{equation}
By equivariance,
\begin{equation}
  \begin{split}
      F(\zeta(t,\x)\sigma(\p)) &= F(\sigma(\p) \zeta(t, \x + \tfrac1m \p t) e^{\frac1m(\x \cdot \p + \frac1{2m}\|\p\|^2)M})\\
      &= e^{-i(\x \cdot \p + (Et + \frac1{2m}\|\p\|^2))} F(\sigma(\p)),
  \end{split}
\end{equation}
and hence, integrating,
\begin{equation}
  \phi(t,\x) = e^{-i E t} \int_{\RR^3} d^3p\ e^{-i(\x \cdot \p + \frac1{2m}\|\p\|^2)} \psi(\p),
\end{equation}
which is up to the $t$-dependent phase in front of the integral,
essentially the Fourier transform of the rescaled function
$e^{-\frac{i}{2m}\|\p\|^2} \psi(\p)$.

As shown in \cite[Appendix~A]{Figueroa-OFarrill:2023qty}, the action
of the Bargmann group on such a field $\phi$ is given by
\begin{equation}
  (g \cdot \phi)(t,\x) = h^{-1} \cdot \phi(t',\x'),
\end{equation}
where $t'$, $\x'$ and $h \in G_0$ are defined by
\begin{equation}
  g^{-1} \zeta(t,\x) = \zeta(t',\x') h.
\end{equation}
Letting $g= \sg(a_+,a_-,\ba,\bv,R)$, we calculate
\begin{equation}
  \begin{split}
    t' &= t - a_-\\
    \x' &= R^{-1}(\x-\ba + (t-a_-)\bv)\\
    h &= \sg(-a_+ - \bv \cdot (\x - \ba) - \tfrac12 (t-a_-)\|\bv\|^2, 0, \bzero, -R^{-1}\bv, R^{-1}),
  \end{split}
\end{equation}
so that
\begin{equation}
  h^{-1} = \sg(a_+ + \bv \cdot (\x - \ba) + \tfrac12 (t-a_-)\|\bv\|^2,  0, \bzero, \bv, R),
\end{equation}
and hence
\begin{equation}
  h^{-1}\cdot \phi(t',\x') =e^{im(a_+ + \bv \cdot (\x - \ba) +
    \tfrac12 (t-a_-)\|\bv\|^2)} R\cdot \phi(t-a_-, R^{-1}(\x- \ba + (t- a_-)\bv )).
\end{equation}
In summary,
\begin{equation}
  \label{eq:G-on-gal-field}
  (g \cdot \phi)(t,\x) =e^{im(a_+ + \bv \cdot (\x - \ba) +
    \tfrac12 (t-a_-)\|\bv\|^2)} R\cdot \phi(t-a_-, R^{-1}(\x- \ba + (t- a_-)\bv )).
\end{equation}
Breaking this transformation into its different components, we find
that
\begin{itemize}
\item under translations,
  \begin{equation}
    (g \cdot \phi)(t,\x) = \phi(t-a_-, \x - \ba);
  \end{equation}
\item under rotations,
  \begin{equation}
    (g \cdot \phi)(t,\x) = R \cdot \phi(t, R^{-1}\x);
  \end{equation}
\item under boosts,
  \begin{equation}
    (g\cdot \phi)(t,\x) = e^{im(\bv \cdot\x + \tfrac12 t\|\bv\|^2)} \phi(t, \x + t\bv ));
  \end{equation}
\item and under the action of the Bargmann central element it
  transforms with a constant phase:
  \begin{equation}
    (g \cdot \phi)(t,\x) =e^{ima_+} \phi(t, \x).
  \end{equation}
\end{itemize}

\subsection{Massless galilean fields}
\label{sec:massl-galil-fields}

Now we will describe the massless UIRs of type $\Romanbar{IV}(n,p)$ as
galilean fields. These are honest (as opposed to projective) UIRs of
the Galilei group. As described in Section~\ref{sec:uirs-orbits-6},
they are carried by square-integrable sections of a complex line
bundle over $\RR \times S^2$ obtained by pulling back the bundle
$\cO(-n)$ over $S^2$. They can be described locally by complex-valued
functions $\psi(s,z)$ on $\RR \times \CC$ with $z$ a stereographic
coordinate on $S^2$. The treatment here is very similar to that of
\cite[Section~4.3]{Figueroa-OFarrill:2023qty} to which we will refer
for the pertinent calculations. The inducing representation is a
complex one-dimensional representation of $K_\tau \ltimes T$, with $T$
acting via the unitary character associated to $\tau = (0,0,\p)$ and
$K_\tau \cong \Spin(\p^\perp) \ltimes \p^\perp$ acting in such a way
that $\p^\perp$ acts trivially and $\Spin(\p^\perp)$ acts with weight
$n \in \ZZ$. To describe the UIR as fields on Galilei spacetime, we
need to embed this one-dimensional representation into an irreducible
(without loss of generality) representation of $K$. We demand that the
boosts act trivially, but must embed the weigh-$n$ representation of
$\Spin(2)$ into an irreducible representation of $\Spin(3)$. As was
done in \cite{Figueroa-OFarrill:2023qty} for the Carroll particles, we
may choose any complex irreducible representation of $\Spin(3)$ of
spin $j \geq |n/2|$. The smallest such representation is that of spin
$j = |n/2$ into which the inducing representation embeds as the
subspace with highest (if $n\geq 0$) or lowest (if $n\leq 0$) weight.
Let us denote by $V$ this complex ($|n|+1$)-dimensional representation
with $T$ acting via the character
\begin{equation}
  \chi_\tau(e^{a_+ M - a_- H + \ba \cdot \bP}) = e^{-i \ba \cdot \p}.
\end{equation}
Let $F: G \to V$ be a $K_\tau \ltimes T$-equivariant Mackey function
and let $\zeta : G/G_0 \to G$ be the coset representative $\zeta(t,\x)
= e^{t H + \x \cdot \bP}$.  Then the galilean field is given by
\begin{equation}
  \phi(t,\x) = \widehat F(\zeta(t,\x)),
\end{equation}
where $\widehat F$ is the group-theoretical Fourier transform
\begin{equation}
  \label{eq:fourier-transform-IV}
  \widehat F(g) = \int_{\RR \times \CC}\frac{2i ds \wedge dz \wedge
    d\zbar}{(1+|z|^2)^2} \sigma(s,z) \cdot F(g \sigma(s,z)),
\end{equation}
where $\sigma(s,z)$ is given by equation~\eqref{eq:coset-rep-IV}.
We calculate
\begin{equation}
  \zeta(t,\x) \sigma(s,z) = e^{t H + \x\cdot\bP} S(z) e^{s/p^2 \p
    \cdot \bB} = \sigma(s,z) e^{t H + (S(z)^{-1}\x - s t/p^2 \p)\cdot \bP},
\end{equation}
where we have ignored terms multiplying $M$ since they act trivially
on massless representations and we can essentially pretend that we are
dealing with the Galilei group, where $\bB$ and $\bP$ commute.  Using
equivariance of $F$ and the fact that $\psi(s,z) = F(\sigma(s,z))$ we
arrive at
\begin{equation}
  \label{eq:gal-fields-IV}
  \phi(t,\x) =\int_{\RR \times \CC} \frac{2i ds \wedge dz \wedge
    d\zbar}{(1+|z|^2)^2} e^{i (ts - \x \cdot \bpi(z))} \rho(S(z)) \psi(s,z),
\end{equation}
where $\rho : \Spin(3) \to \GL(V)$ is the representation
of $\Spin(3)$.

To describe the action of $G$ on such fields, we let $g =
\sg(a_+,a_-,\ba,\bv,R)$ and we calculate
\begin{align}
  g^{-1} \zeta(t,\x) &= R^{-1} e^{-\bv \cdot\bB} e^{-a_+ M + (a_- + t)H + (\x - \ba)\cdot \bP}\\
                     &= \zeta(t+a_-, R^{-1}(\x -\ba - (a_- +t)\bv)) R^{-1} e^{-\bv \cdot \bB},
\end{align}
where we once again have ignored terms in $M$ in the final
calculation.  Using equivariance of the Fourier-transformed Mackey
function \eqref{eq:fourier-transform-IV}, we find that
\begin{align}
  (g \cdot \phi)(t,\x) &= \widehat F(g^{-1}\zeta(t,\x))\\
                       &= R \cdot \phi(t+a_-, R^{-1}(\x - \ba - (t + a_-)\bv)).
\end{align}
Since $\|\bpi(z)\|^2 = p^2$, we may insert zero in the form
$\|\bpi(z)\|^2 - p^2$ in the integrand of
equation~\eqref{eq:gal-fields-IV} and using that any $\bpi(z)$ in the
integrand is the result of differentiating with $i
\boldsymbol{\nabla}$, we see that $\phi(t,\x)$ obeys the Helmholtz
equation
\begin{equation}
  \label{eq:helmholtz-IV}
  (\bigtriangleup +p^2) \phi(t,\x) = 0,
\end{equation}
with $\bigtriangleup$ the laplacian in three-dimensional euclidean
space.  This is the only equation for the inducing representation with
$n=0$, but for $n\neq 0$, we have additional equations.  This is
because the field $\phi$ is $V$-valued and in order to recover the
UIR, we need to project to the inducing one-dimensional
representation, which corresponds to the kernel of $J_+$ (if $n>0$) or
$J_-$ (if $n<0$), where
\begin{equation}
  J_+ =
  \begin{pmatrix}
    0 & 1 \\ 0 & 0
  \end{pmatrix}
  \qquad\text{and}\qquad
  J_+ =
  \begin{pmatrix}
    0 & 0 \\ 1 & 0
  \end{pmatrix}.
\end{equation}
Of course $J_\pm$ live in the complexification of $\so(3)$ and we
extend the representation complex-linearly. We proceed as in
\cite[Section~4.3]{Figueroa-OFarrill:2023qty}.

Let $n>0$ for definiteness and consider
\begin{align}
  0 &= \int_{\RR \times \CC} \frac{2i ds \wedge dz \wedge d\zbar}{(1+|z|^2)^2} e^{i (ts - \x \cdot \bpi(z))} \rho(S(z)) \rho(J_+) \psi(s,z)\\
    &= \int_{\RR \times \CC} \frac{2i ds \wedge dz \wedge d\zbar}{(1+|z|^2)^2} e^{i (ts - \x \cdot \bpi(z))} \rho(J_+(z)) \rho(S(z)) \psi(s,z) \,,
\end{align}
where
\begin{equation}
  J_+(z) = S(z) J_+ S(z)^{-1} = \frac1{1+|z|^2} \begin{pmatrix} -z & z^2 \\ -1 & z \end{pmatrix} \,.
\end{equation}
This is formally the same expression as in
\cite[Section~4.3]{Figueroa-OFarrill:2023qty} and we may borrow the
results from that paper. For helicity $\pm \frac12$ the equations are
the massive Dirac equation in three-dimensional euclidean space:
\begin{equation}
  \label{eq:dirac-like}
  (\slashed{\d} \pm i p) \phi = 0,
\end{equation}
where $\slashed{\d} = \gamma^i \d_i$ with $\gamma^i$ the representation
of $\Cl(0,3)$ given by
\begin{equation}
  \gamma^1=
  \begin{pmatrix}
    0 & -1 \\ -1 & 0
  \end{pmatrix},\qquad
  \gamma^2=
  \begin{pmatrix}
    0 & -i \\ i & 0
  \end{pmatrix} \qquad\text{and}\qquad
  \gamma^3=
  \begin{pmatrix}
    1 & 0 \\ 0 & -1
  \end{pmatrix}.
\end{equation}
Notice that either of these equations imply the Helmholtz
equation~\eqref{eq:helmholtz-IV}.  Similarly, as in
\cite[Section~4.3]{Figueroa-OFarrill:2023qty}, for helicity $\pm1$ we
obtain the field equation for topologically massive Maxwell theory
\cite{Deser:1981wh,Deser:1982vy}
\begin{equation}
  \pm p \phi_i = \epsilon_{ijk} \d_j \phi_k \,,
\end{equation}
which again implies the Helmholtz equation~\eqref{eq:helmholtz-IV}.

\section*{Acknowledgements}

The work of S.~Pekar was supported by the Fonds de la Recherche
Scientifique -- FNRS under grant FC.36447, the SofinaBo\"el Fund for
Education and Talent, the Foundation of the École polytechnique and
the Alexander Friedmann Fund for cosmology. The research of A. Pérez
is partially supported by Fondecyt grants No 1211226, 1220910 and
1230853.

\appendix

\section{Symmetries of the massive spinless Galilei particle}
\label{sec:symm-mass-spinl}

In this appendix we provide further details concerning the symmetries
of the massive spinless Galilei particle in
Section~\ref{sec:mass-galil-part}. The starting point of our analysis
was the following action
\begin{equation}
  \label{eq:Lagr-start-app}
 L[a_{+},t,\x,\bv,R(\bm{\varphi})] = m \dot a_+ - \left(E_{0} + \tfrac12 m
  \|\bv\|^2\right) \dot t  + m \bv \cdot \dot \x \,.
\end{equation}
To obtain the global symmetries we restrict our generic
symmetries~\eqref{eq:symmetries-general} to the representative at
hand, i.e., we set $\p=\bk=\bj=\bzero$ to obtain
\begin{subequations}
  \label{eq:symmetries-general-rest}
  \begin{align}
    \delta_{c_{+}} a_{+} &= c_{+}  &&&&&    m_{Q} &= m \\
    \delta_{c_{t}} t &= c_{t}  &&&&&    E_{Q} &=  \tfrac12 m \|\bv\|^2 + E_{0} \\
    \delta_{c_{x}} \x &= \bm{c}_{x} &&&&& \Rightarrow\quad   \p_{Q} &=  m \bv\\
    \delta_{c_{v}} \bv &=\bm{c}_{v} &\delta_{c_{v}} a_{+} &= -\x \cdot \bm{c}_{v} & \delta_{c_{v}} \x &=t \bm{c}_{v} &    \bk_{Q} &= m \bv t - m \x  \\
    \delta_{\omega} R &= \omega R & \delta_{\omega} \x &= \omega \x & \delta_{\omega} \bv &= \omega \bv  &  \bj_{Q} &=m \x \times \bv \,.
  \end{align}
\end{subequations}
The gauge symmetries are given by the stabiliser of $\alpha$ which for
the massive spinless particles is given by
$\sg(a_+,a_-,\bzero,\bzero,R)$. Infinitesimally they are given by the
following transformations (taken from \eqref{eq:gaugeinv-general})
\begin{subequations}
  \label{eq:gaugeinv-general-app}
  \begin{align}
    \delta_{\lambda_{+}} a_{+}  & = \lambda_{+} \\
    \delta_{\lambda_{t}} t      & = \lambda_{t} & \delta_{\lambda_{t}}a_{+}  & = - \tfrac12 \|\bv\|^2 \lambda_{t} & \delta_{\lambda_{t}}\x & =\bv \lambda_{t}                \\
    \delta_{\lambda_{\omega}} R & = R\lambda_{\omega} \, .
  \end{align}
\end{subequations}
One can explicitly show that they are indeed symmetries (up to
boundary terms) of the action~\eqref{eq:Lagr-start-app}.

The global transformation of $a_{+}$ with parameter $c_+$ are the
time-independent part of a gauge transformation. This piece of the
action can also be written as
$L[p_{+},a_{+},u]=p_{+}\dot a_{+}-u \phi$ where $u$ enforces the
constraint $\phi=p_{+}-m$. This action has no physical degrees of
freedom, but the equations of motion $\dot a_{+}=u$, $\dot p_{+}=0$
and $p_{+}=m$, show that there exists a canonical variable $p_{+}$
that is constant along the trajectory and equal to $m$. Since $p_{+}$
commutes with the first-class constraint it is an observable.

After the canonical analysis we obtained the following action
\begin{equation}
  \label{eq:reparem-2}
  L_{\mathrm{can}}[t,p_{t},\x,\bm{p},u]
  = p_{t}\dot t + \bm{p}\cdot \dot \x - u 
  \left(
   p_{t} + \frac{1}{2m} \| \bm{p}\|^2 + E_{0} 
  \right) ,
\end{equation}
with variation
\begin{equation}
  \label{eq:rep-var}
 \vd  L_{\mathrm{can}}
  = (\dot t - u)\vd p_{t}+ \dot p_{t} \vd t
+ \left(\dot \x -\tfrac{1}{m} u\bm{p}\right) \cdot \vd \x  - \dot{\bm{p}}\cdot \vd \x 
  -
  \left(
   p_{t} + \frac{1}{2m} \| \bm{p}\|^2 + E_{0} 
  \right) \vd u 
  +\frac{d}{d\tau}
  \left(
  p_{t} \vd t + \bm{p}\cdot\vd \x
  \right)
\end{equation}
and global symmetries
\begin{subequations}
  \label{eq:symmetries-repara}
  \begin{align}
\delta_{c_{t}} t &= c_{t}  &&&&&    E_{Q} &=  -p_{t}\approx  \frac{1}{2m} \|\p\|^2 + E_{0}\\
\delta_{c_{x}} \x &= \bm{c}_{x} &&&&& \Rightarrow\quad   \p_{Q} &= \p \\
\delta_{c_{p}} \p &= m \bm{c}_{p} &\delta_{c_{p}} p_{t} &= -\tfrac{1}{m}\p \cdot \delta_{c_{p}} \p & \delta_{c_{p}} \x &=t \bm{c}_{p} &    \bk_{Q} &= t \p- m \x \\
 \delta_{\omega} \x &= \omega \x & \delta_{\omega} \p &= \omega \p  &&&  \bj_{Q} &=\x \times \p \, .
\end{align}
\end{subequations}
The Poisson brackets are given by $\{t,p_{t}\}=1$ and
$\{x_{i},p_{j}\} = \delta_{ij}$ and the gauge transformations
generated by the gauge constraint via
$\delta_{\lambda}F = \lambda \{F,\phi\}$ are given by
\begin{align}
  \label{eq:reparam}
  \delta_{\lambda_{t}} t      & = \lambda_{t} &  \delta_{\lambda_{t}}\x & =\frac{\p}{m} \lambda_{t}  &\delta_{\lambda_{t}}u  & = \dot \lambda_{t} \, .
\end{align}
This is the remaining reparametrisation freedom in $\tau$ and we
accompanied it by a transformation of the Lagrange multiplier $u$ such
that it is a symmetry of the action~\eqref{eq:symmetries-repara}. A
more geometric way to write these gauge transformations is by
transforming all canonical variable as $\vd_{\lambda}z=\dot z \lambda$
and the Lagrange multiplier as
$\vd_{\lambda}u=\frac{d}{d\tau}(u \lambda)$, where we see that the
canonical variables transform as scalars while the $u$ is a scalar
density (see, e.g., Section 4.3.1.\ in \cite{Henneaux:1992ig}).

After gauge fixing the action has the following form
\begin{align}
  \label{eq:massive-can-final-app}
  L_{\mathrm{can}}[\x,\bm{p}]=\bm{p}\cdot \dot \x  -
  \left(
  \frac{1}{2m} \| \bm{p}\|^2 +E_{0}  
  \right) ,
\end{align}
where the Hamiltonian is given by
$E=\frac{1}{2m} \| \bm{p}\|^2 +E_{0}$ and symmetries are now given by
\begin{subequations}
  \label{eq:symmetries-repara-2}
  \begin{align}
\delta_{c_{x}} \x &= \bm{c}_{x} &&&    \p_{Q} &= \p \\
\delta_{c_{p}} \p &= m \bm{c}_{p} & \delta_{c_{p}} \x &=t \bm{c}_{p} &  \Rightarrow\quad  \bk_{Q} &= t \p- m \x \\
 \delta_{\omega} \x &= \omega \x & \delta_{\omega} \p &= \omega \p  &  \bj_{Q} &=\x \times \p \, .
\end{align}
\end{subequations}

We used $\p= m \dot\x$ to write the action in configuration space
\begin{align}
  \label{eq:massive-conf-app}
  L_{\mathrm{red}}[\x]=  \frac{m}{2} \| \dot \x\|^2 -  E_{0}  \, ,
\end{align}
where it has the following symmetries
\begin{subequations}
  \label{eq:symmetries-conf}
  \begin{align}
\delta_{c_{x}} \x &= \bm{c}_{x} &&&    \p_{Q} &= m \dot\x \\
 \delta_{c_{p}} \x &=t \bm{c}_{p} &&&  \Rightarrow\quad  \bk_{Q} &= t m \dot\x- m \x \\
 \delta_{\omega} \x &= \omega \x & \delta_{\omega} \p &= \omega \p  &  \bj_{Q} &= \x \times  m \dot\x \, .
\end{align}
\end{subequations}

\providecommand{\href}[2]{#2}\begingroup\raggedright\endgroup

\end{document}